\documentclass{article}

\usepackage{microtype}
\usepackage{graphicx}
\usepackage{subfigure}
\usepackage{booktabs} 

\usepackage{hyperref}

\usepackage{balance} 

\usepackage[accepted]{icml2025}

\usepackage{amsmath}
\usepackage{amssymb}
\usepackage{mathtools}
\usepackage{amsthm}

\usepackage[capitalize,noabbrev]{cleveref}

\theoremstyle{plain}

\theoremstyle{definition}

\theoremstyle{remark}

\usepackage[textsize=tiny]{todonotes}

\usepackage{multirow}
\usepackage{dsfont}

\usepackage{balance} 

\usepackage{xcolor}
\usepackage[font=small,skip=3pt]{caption}
\newcommand\sk[1]{{\color{black}{#1}}}

\icmltitlerunning{Evaluating and Improving Graph-based Explanation Methods for Multi-Agent Coordination}

\begin{document}

\twocolumn[
\icmltitle{Evaluating and Improving Graph-based Explanation Methods \\ for Multi-Agent Coordination}



\icmlsetsymbol{equal}{*}

\begin{icmlauthorlist}
\icmlauthor{Siva Kailas}{yyy}
\icmlauthor{Shalin Jain}{yyy}
\icmlauthor{Harish Ravichandar}{yyy}
\end{icmlauthorlist}

\icmlaffiliation{yyy}{School of Interactive Computing, College of Computing, Georgia Institute of Technology, Atlanta, Georgia, United States of America}

\icmlcorrespondingauthor{Siva Kailas}{skailas3@gatech.edu}

\icmlkeywords{Machine Learning, ICML}

\vskip 0.3in
]



\printAffiliationsAndNotice{}  

\begin{abstract}
Graph Neural Networks (GNNs), developed by the graph learning community, have been adopted and shown to be highly effective in multi-robot and multi-agent learning. 
Inspired by this successful cross-pollination, we investigate and characterize the suitability of existing GNN explanation methods for explaining multi-agent coordination.
We find that these methods have the potential to identify the most-influential communication channels that impact the team's behavior. 
Informed by our initial analyses, we propose an attention entropy regularization term that renders GAT-based policies more amenable to existing graph-based explainers. 
Intuitively, minimizing attention entropy incentivizes agents to limit their attention to the most influential or impactful agents, thereby easing the challenge faced by the explainer. 
We theoretically ground this intuition by showing that minimizing attention entropy increases the disparity between the explainer-generated subgraph and its complement. 
Evaluations across three tasks and three team sizes i) provides insights into the effectiveness of existing explainers, and ii) demonstrates that our proposed regularization consistently improves explanation quality without sacrificing task performance. 
\end{abstract}

\section{Introduction}
\label{related:GNN_MAC}

Graph neural networks (GNNs) were originally developed to analyze complex relational data~\cite{wu2020comprehensive}. However, they were quickly adopted by various other communities due to their ability to capture structural information and reason over non-euclidean spaces while remaining invariant to certain distractors. The fields of multi-agent and multi-robot learning were among the beneficiaries of these powerful techniques, enabling scalable policies that encode team size-invariant strategies for inter-robot communication and coordination. Indeed, researchers have demonstrated that GNNs are effective in solving a variety of core challenges in multi-agent coordination \cite{coordination-graphs}, such as information aggregation~\cite{nayak2023scalable}, decentralization~\cite{ji2021decentralized}, and learning to communicate~\cite{sheng2022learning}.
This has instantiated into adoption of RL trained GNN-based policies with overall similar design choices in the multi-robot community to tackle practical applications such as cooperative navigation \cite{porok-path-plan, magat}, coverage control \cite{coverage-control}, autonomous driving \cite{gat-auto-drive}, and real-world multi-robot coordination \cite{passageProrok}.

\begin{figure}
    \centering
    \includegraphics[width=\linewidth]{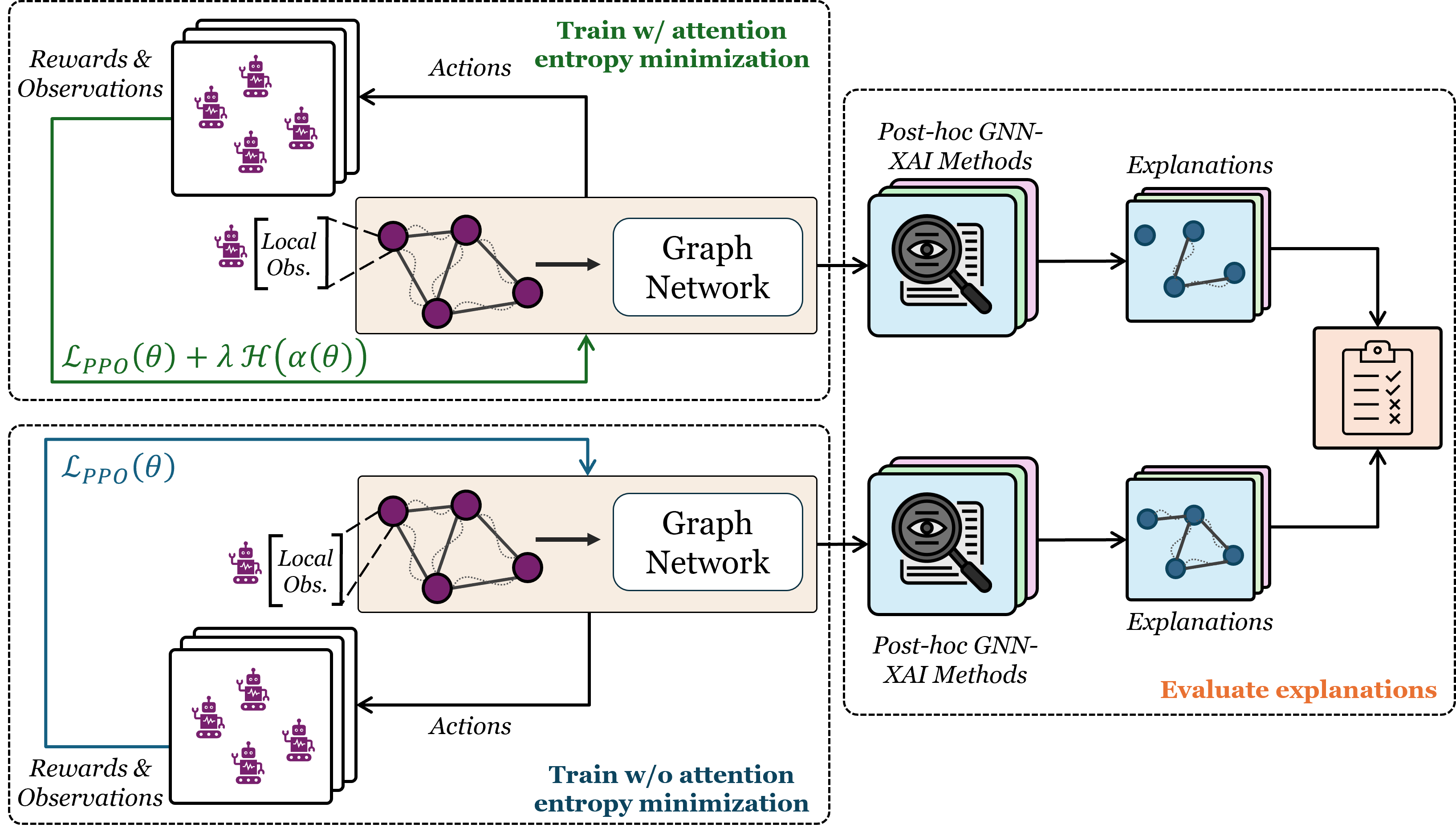}
    \caption{We systematically evaluate existing graph-based explainers when used to explain multi-agent coordination, and propose a regularizer to improve the explainability of GNN-based policies.}
    \label{fig:blockdiagram}
\end{figure}

Despite the impressive strides in multi-agent learning fueled by GNNs, the inner workings of the learned policies remain opaque to most users. GNN-based coordination policies are hard to introspect and analyze, giving rise to unexplained agent behaviors and decision making strategies. In contrast, other fields that analyze and study graph-structured data (e.g., material science~\cite{reiser2022graph} and chemistry~\cite{yang2021predicting}) enjoy improved transparency and explainability, thanks to recent developments in graph-based explanation methods~\cite{gnnexplainer}.
These aforementioned works, and the graph learning community in general, have been pursuing explanations of GNNs with respect to supervised learning, and typically do so on large graph data. Thus, analogous to how the multi-agent/multi-robot community has evaluated and transposed the use of GNNs from the graph learning community for learning coordination for different tasks (as evidenced by the aforementioned related works), we seek to do a similar evaluation with GNN-based post-hoc explainers. 
This evaluation, combined with the insights presented to potentially improve the explanation quality, can be used to address a much-needed explanation paradigm for multi-agent coordination (see \cite{brandao2022explainability}.
and \sectionautorefname~\ref{sec:related_works}).

In this work, we explore whether one could adopt existing GNN explanation methods to explain multi-agent coordination. We systematically investigate and characterize the suitability of existing GNN explanation methods for graph attention network (GAT) based policies in multi-agent coordination. If we could explain GNN-based coordination policies, users can effectively debug the learning algorithm by comparing observed coordination strategies against their expectations. Further, explanations would help non-experts gain insights into learned coordination policies. 

We study GNN explainers for multi-agent coordination since they estimate parsimonious yet representative subgraphs as a means to explain complex decision making over graphs. 
When translated to multi-agent coordination, extracting such subgraphs can help identify the most influential inter-agent interactions that can effectively approximate and distill coordination strategies learned across the entire team. 
In fact, identifying such influential interactions was found to be a key challenge in explaining coordination strategies by a recent user study focused on multi-agent navigation~\cite{brandao2022explainability}. Further, these explanation methods are agnostic to the learning algorithm or paradigm used to train GNNs and do not interfere with training or task performance.


In particular, we analyzed three prominent post-hoc graph-based explainers (Graph Mask~\cite{graphmask}, GNN-Explainer~\cite{gnnexplainer} and Attention Explainer~\cite{Fey/Lenssen/2019}) across three multi-agent tasks (blind navigation, constrained navigation, and search and rescue) and measured the quality of their explanations using established explanation metrics (e.g., fidelity~\cite{amara2022graphframex} and faithfulness~\cite{agarwal2023evaluating}). 

Our findings suggest that graph-based explainers have the potential to explain learned inter-agent influences in GNN-based coordination policies. 
However, our analysis also revealed that there is room for improvement in the quality of explanations generated by these approaches.

In addition to our systematic analyses, we propose a simple regularization technique for training graph-based coordination policies in an effort to make learned policies more amenable to existing explanation methods. Specifically, we introduce an attention entropy minimization objective that can be used as a regularizer when training graph attention networks (GATs). Intuitively, minimizing attention entropy will ``narrow" each agent's attention to the most influential or impactful neighbors. In turn, this more focused attention simplifies the core problem faced by graph-based explainers: identifying the most-influential inter-agent interactions.

Our theoretical analysis shows that minimizing attention entropy increases the disparity between the subgraph generated based on attention values and its complement. This aligns well with intuition and helps explain the proposed regularizer's potential to improve explanation quality.

Rigorous empirical evaluations across three tasks and differing team sizes suggest that our regularization approach is remarkably effective on Attention Explainer, transforming a simple explanation method into the best-performing method in terms of explanation quality across all tasks.
We find that pairing minimizing attention entropy with the other two explainers tends to improve at least one aspect of their explanation quality.
Further, we find our regularization's improvements to explanation quality come with negligible impact on task performance. 
Our theoretical and empirical analyses provide a strong foundation and take the first steps towards interpretable and explainable graph-based policies for multi-agent coordination.

In summary, our contributions include: i) insights from a systematic evaluation of existing post-hoc graph-based explainers when used to explain GNN-based multi-agent coordination, and ii) a theoretically-grounded attention entropy regularization scheme that is shown to improve the explainability of learned GNN-based policies.


\section{Related Works}
\label{sec:related_works}

\textbf{Explainable multi-agent coordination}:
Since explainable multi-agent coordination was proposed as a new research direction recently \cite{kraus2020ai}, only a few works have explored this problem. 
One line of work investigates explanations via policy abstraction using a multi-agent MDP~\cite{boggess2022toward, boggess2023explainable}. Similarly, policy summarization and the use of landmarks to condition and convey the high-level strategy is also being explored~\cite{pandya2024multi}. These works rely on text and visual modalities, and are focused more on high-level explanations. As a result, they do not focus on explaining or distilling key interactions among agents.
\sk{\textbf{In fact, explanations that capture critical and affected agents (and agent-agent influences) were found to be desirable in a user-study by \cite{brandao2022explainability} for multi-agent/multi-robot navigation-based tasks.}}
While a different line of work has proposed generating easily verifiable multi-agent path finding plans as explainable~\cite{almagor2020explainable, kottinger2022conflict, kottingerexplainable}, they inherently do not capture key agents or interactions as desired and are limited to specific problem representations of multi-agent navigation. These gaps in the literature strongly motivate the idea of identifying subgraphs of agents to outline the most relevant agents and their interactions among one another. Toward this, we investigate and improve the utility of graph-based explanations of GNN-based multi-agent policies.

\textbf{Graph-based explanation methods}:
The prevalent approach to explaining GNNs is to identify the subgraph most relevant to its prediction~\cite{graphmask, gnnexplainer, luo2020parameterized, yuan2022explainability}. In addition to these methods, other approaches include generating probabilistic graphical models to explain GNN predictions \cite{vu2020pgm}, identifying graph patterns and motifs \cite{yuan2020xgnn}, and generating counterfactual subgraphs representing the minimal perturbation such that the GNN prediction changes \cite{lucic2022cf}. These methods have demonstrated success in explaining the predictions of GNNs on synthetic graph datasets such as Tree-Cycles and BA-shapes \cite{gnnexplainer}, and large real-world graph datasets such as MUTAG \cite{debnath1991structure}. However, their ability to explain graph-based multi-agent policies remains unexplored. Motivated by the potential of subgraph explanations to identify the most influential channels of communication within a team, we focus on subgraph identification methods. To explore differing subgraph explanations, we use GraphMask to obtain binary mask subgraphs \cite{graphmask}, GNN-Explainer to obtain continuous mask subgraphs \cite{gnnexplainer}, and Attention-Explainer to obtain attention-based subgraphs for Graph Attention Networks \cite{Fey/Lenssen/2019}.


\section{Preliminaries}

\textbf{Multi-Agent Coordination}:
We consider a class of multi-agent coordination problems that can be modeled as Decentralized Partially Observable Markov Decision Processes (POMDP). They are described by the tuple $(\mathcal{D, S, A, T, R, O})$, where $\mathcal{D}$ is the set of N agents, $\mathcal{S}$ is the set global states, $\mathcal{A}$ is the set of actions each agent can take, $\mathcal{T}$ is the state transition model, $\mathcal{R}$ is the global reward function, and $\mathcal{O}$ is the observation model. We aim to learn decentralized action policies $\pi(a_i|o_{\mathcal{N}(i)})$ that maximize the expected return $\mathds{E}[\sum^T_{t=0} r_t]$, where each agent's policy is conditioned its own observations and the messages received from its neighbors through a graph network. We include the ego agent in its own set of neighbors.

\textbf{Graph Neural Networks}:
We consider policies that contain a convolutional Graph Neural Network (GNN). GNNs consist of $L$ layers of graph convolutions, followed by a nonlinear activation~\cite{gcn}. GNNs facilitate learned communication between agents over the communication graph $G\mathcal{(V, E)}$, where $\mathcal{V} = \{v_1, ... v_N\}$ is the set of $N$ nodes representing agents and $(v_i, v_j) \in \mathcal{E}$ indicates a communication link between agents $i$ and $j$. A single Graph Convolution Layer is given by $h_i^{'} = \sum_{j \in \mathcal{N}(i)} \frac{1}{\sqrt{deg(i)deg(j)}}\phi_\theta h_j$
where $h_j$ is the node feature of $v_j$, $\phi_\theta$ is a learned transformation of the node feature parameterized by $\theta$, and $\mathcal{N}(i) = \{j~|~(v_i, v_j) \in \mathcal{E}\}$.
\sk{Early works in multi-robot coordination using GNNs adopted this base architecture (e.g., \cite{porok-path-plan}). However, this architecture implicitly assumes that all agents exhibit equivalent influence through the message passing/communication channels.}
\sk{
This is generally not the case, and more recent works have pivoted to adopting graph attention layers into their architecture in lieu of standard GNN layers due to the improved expressivity and performance achieved in multi-agent/multi-robot tasks (e.g., \cite{magat, gat-auto-drive, he2023multi}).
In this work, every policy contains a single Graph Attention Layer \cite{gatv2}, which extends the graph convolution layer with a learned attention function that weights each edge $(v_i, v_j) \in \mathcal{E}$. A single graph attention layer is given by $h_i^{'} = \sum_{j \in \mathcal{N}(i)} \alpha_{ij} \cdot \phi_\theta h_j$, where $\alpha_{ij}$ is the edge weight computed by the attention function introduced by \citet{gatv2}.
}

\textbf{GNN-based Multi-Agent Coordination}:
There are fundamental design choices that have been adopted throughout the GNN-based multi-agent coordination literature (see Sec. ~\ref{sec:related_works}). Most importantly, an agent-agent graph and a representation of each agent's local observations are provided as inputs to the GNN as the adjacency matrix and the node features. Within the GNN, single or several rounds of message passing are conducted through graph convolutional layers, allowing agents to integrate information from surrounding agents into their own node representation (such as local observations), analogous to communication channels. This node representation is then utilized for task completion, typically via a local decoder structure such as a multi-layer perceptron with shared parameters/weights. This effectively conditions the action of the ego agent on the observations of surrounding agents in addition to the ego agent. We adopt these fundamental design aspects in our model architecture, along with task-specific design choices informed by prior works~\cite{bettini2024benchmarl, bettini2022vmas, passageProrok}.


\section{Evaluating Graph Explainers}
Below, we discuss the graph explainers we investigate along with the tasks and metrics we use to evaluate both explanation quality and task performance. 

\subsection{Graph-based Explainers}
\label{subsec:explainers}
We investigate the following GNN explainers in terms of their ability to explain GNN-based coordination policies.

\textbf{Graph Mask:} Graph Mask \cite{graphmask} searches for the minimal-edge subgraph $G_s \subseteq G$ such that a measure of the divergence between the output under the subgraph and the original graph is below a threshold: $D(f(G_s, \mathcal{X} || f(G, \mathcal{X})) < \beta$. The minimal-edge subgraph $G_s$ is then a binary-valued mask over the original graph.

\label{gnnExplainerPrelim}
\textbf{GNN-Explainer:} GNN-Explainer \cite{gnnexplainer} explains GNNs by searching for the subgraph $G_s \subseteq G$ and node feature subset $\mathcal{X}_s \subseteq \mathcal{X}$ having the maximal mutual information with the prediction under the original graph and node feature set.
In this work, we are most interested in the subgraph identification component, which is a continuous-valued edge mask over the original graph, incentivized to be discrete through regularization.

\textbf{Attention Explainer:} Given the prominence of graph attention networks in policies for multi-agent coordination \cite{magat, gat-auto-drive, collab-perception}, we also investigate Attention Explainer~\cite{Fey/Lenssen/2019} which presents the attention values generated by the Graph Attention Layer as the explanation. Intuitively, these values represent the importance of each channel within the communication graph, with a higher attention value indicating greater importance and influence.

\subsection{Tasks}
We evaluate the three graph-based explainers introduced above on three multi-agent coordination tasks implemented using the VMAS simulator~\cite{bettini2022vmas} across three team sizes.
\sk{
These tasks are all the same or harder versions of the tasks in BenchMARL~\cite{bettini2024benchmarl}, a new MARL benchmark. Since most tasks in BenchMARL and VMAS are unsolved with GNNs (or in general), we adopted the model architecture and training parameters of prior works for tasks that have been solved using GNNs, and designed GNNs for other tasks based on related work.
}

\textbf{Blind Navigation:} We adapted this task from \cite{bettini2022vmas}. The original task considers a team of $N$ agents, each equipped with LIDAR and required to navigate to its assigned goal location while minimizing collisions with other agents. Both initial and goal locations are randomized.
Our \textit{blind} navigation task differs from the original version in that there are no LIDAR sensors equipped on the agents. This means that the agents only observe their own position, velocity, and goal location, and that each agent has no information about the other agents from local observations alone. Thus, effective communication is required to solve our blind version of this task. We consider three team sizes ($N={3,4,5}$).
For all team sizes, we design a common policy with an Identity layer for the encoder, a 32-dim GATv2 layer with TanH activation, and a 2-layer decoder with 256 units each and TanH activation. 
\sk{
We use the dense reward structure from \cite{bettini2022vmas,bettini2024benchmarl}.
}

\textbf{Blind Passage:} We adapted this task from \cite{passageProrok, bettini2022vmas}. The original passage task considers a team of 5 agents that start in a cross-shaped formation on one side of a wall and need to reach a destination on the other side with the same formation after traversing a narrow corridor or passage and minimizing collisions with each other and with the wall.
The original passage task is hard-coded to 5 agents. We modify the environment to also consider 3- and 4-agent teams by having the 3 or 4 agents randomly spawn in any of the 5 positions in the cross-shaped formation. We use the same reward structure and policy architecture (a 4 layer encoder with 32 units each and ReLU activation, a 32-dim GATv2 layer with TanH activation, and a 4 layer decoder with 64 units each and ReLU activation) as the original version, and make no modifications across team sizes.
\sk{
We use the dense reward structure from \cite{passageProrok}.
}

\textbf{Blind Discovery:} This task is a specific instantiation of the Discovery task proposed in \cite{bettini2022vmas}, with modifications applied to the observation space. The task considers a team of $N$ agents, which are spawned randomly and must explore the environment to discover a single landmark.
The landmark is considered discovered when at least 2 agents converge on its position simultaneously. 
We modify the agent observation space with respect to the landmark to have a sensing radius instead of a simulated 12-point LIDAR for training efficiency. Agents can only observe their own position, velocity, and a binary value indicating if the landmark is within their sensing radius. For all team sizes, we design a common policy with a 1-layer encoder with 64 units and a ReLU activation, 64 GATv2 layer with ReLU activation, and a 2-layer decoder with 64 units each and a ReLU activation. 
\sk{
We use the sparse reward structure from \cite{bettini2022vmas}.
}

\subsection{Metrics for Explanation Quality}

We consider four metrics to quantify both the fidelity~\cite{amara2022graphframex} and faithfulness~\cite{agarwal2023evaluating} of generated explanations.
These metrics and their variants have been studied and used extensively in the graph learning community to characterize the quality of subgraph-based explanations~\cite{kosangnnx,liu2021dig,Fey/Lenssen/2019, pope2019explainability, bajaj2021robust, yuan2022explainability}. 
In the definitions below, $F(\cdot)$ refers to the GNN-based model and $G^t = (X^t,A^t)$ refers to the original graph input at timestep $t$, where $A^t$ is the adjacency matrix at timestep $t$ of $G^t$ and $X^t$ is set of all node features at timestep $t$ in $G^t$. We use $G_S^t$ to refer to the explanation subgraph at timestep $t$ with $A_S^t$ as its adjacency matrix at timestep $t$. Since $G^t$ is a fully connected graph at every timestep in our experiments, $A^t = \mathbf{1},\,\forall t$ where $\mathbf{1}$ is a matrix with each entry equal to $1$. 
We define each of explanation quality metric for a single timestep, and we average across all timesteps and across multiple episodes, each with a random initial state, during the evaluation in \sectionautorefname~\ref{empiricalResultsDiscussion}.

\textbf{Positive Fidelity $(\uparrow)$}: 
$Fid_+$ is defined as the difference between the original prediction $F(G)$ and the prediction generated using the complement of the explanation subgraph $F(G \setminus G_S)$ \cite{pope2019explainability, bajaj2021robust}. The adjacency matrix of the complement of the subgraph $G \setminus G_S$ is $A - A_S = \mathbf{1} - A_S$. Formally, $Fid_+$ at timestep $t$ is $Fid_+^t \triangleq |F(G^t) - F(G \setminus G_S^t)|$,
and measures the \textit{necessity} of the explanation subgraph. The more necessary the subgraph, the larger the change to the prediction when it's removed~\cite{amara2022graphframex}. Larger $Fid_+$ are preferred.

\textbf{Negative Fidelity $(\downarrow)$}:
$Fid_-$ is defined as the difference between the original prediction $F(G)$ and the prediction under the explanation subgraph $F(G_S)$ \cite{yuan2022explainability}. Formally, $Fid_-$ at timestep $t$ is $Fid_-^t \triangleq |F(G^t) - F(G_S^t)|$,
and measures the \textit{sufficiency} of the explanation subgraph. The more sufficient the explanation subgraph, the closer its prediction to that of the initial graph~\cite{amara2022graphframex}. Smaller $Fid_-^t$ are preferred.

\textbf{Delta Fidelity  $(\uparrow)$}:
We define $Fid_{\Delta}$ as the difference between positive fidelity ($Fid_+$) and negative fidelity ($Fid_-$). Formally, $Fid_\Delta$ at timestep $t$ is defined as $Fid_{\Delta} \triangleq Fid_+ - Fid_-$. Since we desire an explanation subgraph that results in a large $Fid_+$ and a small $Fid_-$, a high fidelity explanation subgraph $G_S$ should maximize $Fid_{\Delta}$.

\textbf{Unfaithfulness $(\downarrow)$}:
$GEF$ measures how unfaithful the subgraph explanations are to an underlying GNN model \cite{agarwal2023evaluating}. Formally, $GEF$ at timestep $t$ is $GEF^t \triangleq  1 - \exp{(-KL(F(G^t) || F(G_S^t)))}$, 
where $KL(\cdot || \cdot)$ is the Kullback–Leibler divergence. $GEF^t$ is 1 when $KL(F(G^t) || F(G_S^t))$ tends to $\infty$ and is 0 when $KL(F(G^t) || F(G_S^t))$ tends to $0$. As a result, a smaller unfaithfulness measure is
preferred.

\section{Improving Explanations}
\label{subsec:entropy_reg}

After analyzing the quality of explanations generated by existing explanation results (see Sec. \ref{empiricalResultsDiscussion}), we observed that they could be improved by modifying how we train GNNs -- by minimizing the entropy of the attention values.

Attention entropy minimization can be motivated from two perspectives, one from a multi-agent coordination perspective and one from a graph learning perspective. From a multi-agent collaboration perspective, an agent who collaborates with other agents will intuitively desire to (1) filter out useless information and (2) focus on the information that is most crucial to the task at hand. This is akin to how humans use selective attention to filter unnecessary information and focus on the truly important information for decision-making~\cite{moerel2024selective}. Attention in GNNs serve a similar purpose when the nodes are agents and the edges are communication channels composed of agents providing information to other agents. Minimizing attention entropy in a multi-agent setting yields an attention distribution that is further from pure uniform attention, resulting in a set of agents that are given more focus than other agents.

From a graph learning perspective, minimizing attention entropy in conjunction with task objective can be seen as a form of denoising for node-level learning tasks (e.g., node classification or regression). By incentivizing a set of attention values that have lower entropy, the model is likely to learn a stronger filter that starts removing extraneous or noisy information.
The intuition behind minimizing attention entropy can be connected to more formal notions of information bottleneck over graphs \cite{yugraph}. But, optimizing such objectives tends to be computationally intensive and challenging, which will be likely exacerbated when combined with multi-agent learning. In contrast, integrating attention entropy minimization into learning GNNs is much simpler, especially within MARL frameworks like MAPPO \cite{yu2022surprising}. Formally, let $\mathcal{L}^{PPO}_t(\theta)$ be the standard clip PPO loss used in MAPPO \cite{yu2022surprising, schulman2017proximal} to update the policy weights $\theta$ and let $\alpha_t(\theta)$ be the attention values generated from the model parameters $\theta$. We can define the new regularized loss as


\begin{equation}
    \mathcal{L}^{PPO + ATTN}_t = \mathds{E}_t [\mathcal{L}^{PPO}_t(\theta) + \lambda \mathcal{H}(\alpha_t(\theta))]
    \label{eqn:attnEntPPO}
\end{equation}

where $\mathcal{H}(p) = -\sum_i p_i \log (p_i)$ denotes the entropy of $p$. 



\subsection{Theoretical Analysis}


Let $G$ be the fully-connected input graph of $N$ nodes or agents (with $A = \mathbf{1}$ as its adjacency matrix) to a GNN-based model that learns a set of attention values $\alpha_{ij}$ for each pair of vertices $i, j$ in $G$. Moreover, let $G_S(\alpha)$ be the subgraph generated by Attention Explainer using an attention-based adjacency matrix $A_S(\alpha)$ whose $ij$th element is given by $\alpha_{ij}$.
The complement of $G_S(\alpha)$ is $G \setminus G_S(\alpha)$ with the corresponding adjacency matrix $\mathbf{1} - A_S(\alpha)$. Below, we show that minimizing the attention entropy corresponds to maximizing the dissimilarity of the subgraph $G_S$ and its complement $G \setminus G_S$ under a global measure characterized by an entry-wise matrix norm. 

Distances between two graphs $G_1$ and $G_2$ are best described in terms of the adjacency matrix $A_{G_1 - G_2} = A_{G_1} - A_{G_2}$, and mismatch norms are matrix norms applied to $A_{G_1 - G_2}$~\cite{gervens2022graph}. It turns out that global norms such as edit distance and its direct generalizations correspond to entrywise matrix norms, and that the entry-wise matrix $1$-norm $||A||_1 = (\sum_{i, j} |A_{i,j}||)$ is a suitable mismatch norm~\cite{gervens2022graph}. As such, the distance between $G_S(\alpha)$ and $G \setminus G_S(\alpha)$ (denoted as $D(\alpha)$) 
can be represented as the entry-wise 1-norm of $A_S(\alpha) - (\mathbf{1} - A_S(\alpha))$, yielding 
{\scriptsize
\begin{align}
    D(\alpha) &= \left|\left|A_S(\alpha)_{i, j} - (\mathbf{1} - A_S(\alpha)_{i, j})\right|\right|_1\\ 
    &= \sum_{i, j} \left|A_S(\alpha)_{i,j} - (\mathbf{1} - A_S(\alpha)_{i, j})\right| \\
    &= \sum_{i, j} \left|\alpha_{i,j} - (1 - \alpha_{i, j})\right| = \sum_{i, j} \left| 2\alpha_{i,j} - 1 \right|
    \label{eqn:reductionStep}
\end{align}
}
Note that $D(\alpha)$ is convex as it is a sum of convex functions. As such, we can compute its lower as shown below
using Jensen's inequality~\cite{mcshane1937jensen} and the fact that $\sum_j \alpha_{i, j} = 1,\, \forall i$ due to softmax normalization. 
\begin{equation*}
    \scriptstyle
    D(\alpha) = \sum_{i, j} |2\alpha_{i,j} - 1| \geq N^2 \left| 2\cdot\frac{\sum_{i, j}\alpha_{i, j}}{N^2}-1 \right| = N^2 \left| 2\cdot\frac{N}{N^2}-1 \right|
    = \left|2N - N^2\right|
\end{equation*}

Now, suppose $\alpha_{i, j} = \frac{1}{N}, \, \forall i, j$ (that is, a uniform attention distribution is learned for all nodes in $G$). We can determine 

\vspace{-1.5em}
\begin{equation}
    \scriptstyle
    D \left(\alpha = \left[\frac{1}{N}, \ldots, \frac{1}{N}\right] \right) = \sum_{i, j} \left |2 \cdot \frac{1}{N} - 1 \right| 
    \ = N^2 \left|\frac{2}{N} - 1 \right|
    \ = \left|2N - N^2\right|
    \label{eqn:lowerBoundDalpha}
\end{equation}
\vspace{-1.5em}

From the above, we can see that a uniform attention distribution realizes the lower bound of $D(\alpha)$, implying that a uniform attention distribution yields the lowest distance between the induced subgraph explanation $G_S(\alpha)$ and the complement of the subgraph $G \setminus G_S(\alpha)$ (i.e., a global minima under simplex constraint on $\alpha_{i, j=1:N}$). We can additionally analyze what happens to $D(\alpha)$ at the extreme points of the simplex constraint on $\alpha_{i, j=1:N}$ (w.l.o.g., let $\boldsymbol{\alpha} = \alpha_{i, j=1:N} = \left[1, 0, \ldots 0\right], \, \forall i$) as follows

\vspace{-1.75em}

\begin{equation}
\scriptstyle
    D \left(\boldsymbol{\alpha}\ =\ \left[1, 0, \ldots, 0\right], \, \forall i \right) \ =\ \sum_{i, j} 1 \ =\ N^2 \ > \  \left|2N - N^2\right|, \, \forall N \geq 2
    \label{eqn:extremePointEval}
\end{equation}

\vspace{-0.5em}

From this, we can see that the minima is not persistent throughout the simplex on $\alpha_{i, j=1:N}$. Note that through a nearly-identical analysis of negative attention entropy $-\mathcal{H}(\alpha)$, the uniform distribution achieves a global minima under simplex constraint on $\alpha_{i, j=1:N}$ for $-\mathcal{H}(\alpha_{i, j})$ (i.e., the uniform distribution of attention results in the largest entropy). Moreover, as $-\mathcal{H}$ is \textit{strictly} convex, this minima is unique to the uniform distribution. 

Given that $D(\alpha)$ and $-\mathcal{H}(\alpha)$ are both convex functions that share a global minima and that minima is not persistent over the simplex constraint on $\alpha_{i, j=1:N}$, we can conclude that minimizing attention entropy widens the separation between $G_S(\alpha)$ and $G \setminus G_S(\alpha)$. 

We hypothesize that increasing the distance $D(\alpha)$ is a desirable trait to have within the GNN-based policy structure for explainability.
\sk{
Intuitively, increasing $D(\alpha)$ increases the correlation of the original model output $F(G)$ and the model output using the subgraph induced by the attention values $F(G_S(\alpha))$. This can be attributed to attention entropy minimization pruning noisy and detractive edges when jointly guided by the task reward. This improved alignment should manifest as an improvement in negative fidelity, as supported by empirical results.

Conversely, increasing $D(\alpha)$ decreases the correlation of the original model output $F(G)$ and the model output using the complement of the subgraph induced by the attention values $F(G \setminus G_S(\alpha))$. Since the distance $D(\alpha)$ between $G_S(\alpha)$ and $G \setminus G_S(\alpha)$ is maximized under the regularization, coupled with the fact that $G \setminus G_S(\alpha)$ will likely have the noisy and detractive edges as mentioned earlier, it is likely that $F(G)$ and $F(G \setminus G_S(\alpha))$ will diverge, improving positive fidelity (also supported by our empirical experiments).
In effect, the derived relationship between $D(\alpha)$ and attention entropy minimization (i.e. minimizing attention entropy $\rightarrow$ maximizing $D(\alpha)$) suggests that our regularization likely incentivizes the attention-induced subgraph to be both necessary and sufficient via the above mechanisms.
}

\sk{
To summarize, since negative fidelity is proportional to how much the subgraph explanation approximates the original graph's predictions, we will likely see a reduction in negative fidelity when $D(\alpha)$ increases.
Similarly, since positively fidelity is proportional to how much useful information is taken away when the subgraph explanation is deleted from the original graph, we will likely see an improvement in positive fidelity when $D(\alpha)$ increases.
}

Finally, upon inspection of the objective $D(\alpha)$, we find that the gradient within an $\epsilon$-neighborhood of the uniform distribution (i.e., the minima) has a gradient of 0. As such, unless the gradient produced by the learning objective guides the parameters to leave this neighborhood, the distance between $G_S$ and $G \setminus G_S$ will not increase. Under the assumption that this property is desirable (which we confirm in our empirical analysis), a surrogate objective such as entropy $\mathcal{H}$ will provide a better gradient signal due to its strict concavity.

\section{Empirical Results and Discussion}
\label{empiricalResultsDiscussion}


Our empirical evaluation has a twofold purpose. First, to characterize and compare the performance of state-of-the-art graph-based explainers when deployed in cooperative MARL settings as opposed to large graph data and supervised learning settings as done in prior work.
Second, to investigate how minimizing attention entropy within GAT-based policies affects both explanation quality and task performance. For all evaluations, we run 50 rollouts in each environment for each independent variable of interest.
All relative comparisons made below were validated via Mann-Whitney U-tests with Bonferroni correction.

\begin{figure}[t]
    \centering
    \includegraphics[trim={0.1cm 0.65cm 1.55cm 0.8cm},clip,width=0.325\linewidth]{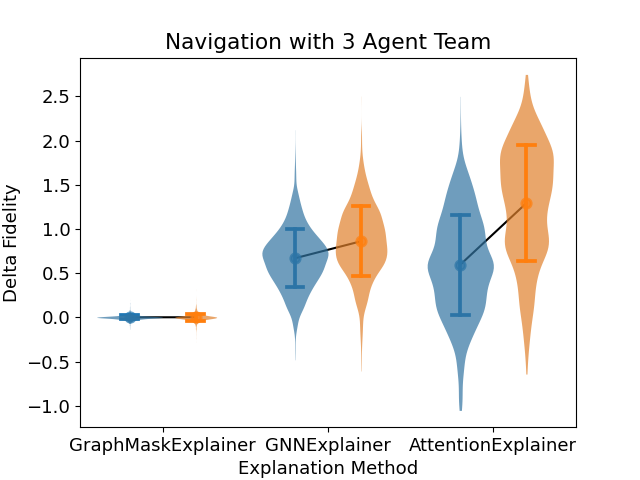}
    \includegraphics[trim={0.5cm 0.65cm 1.55cm 0.8cm},clip,width=0.317\linewidth]{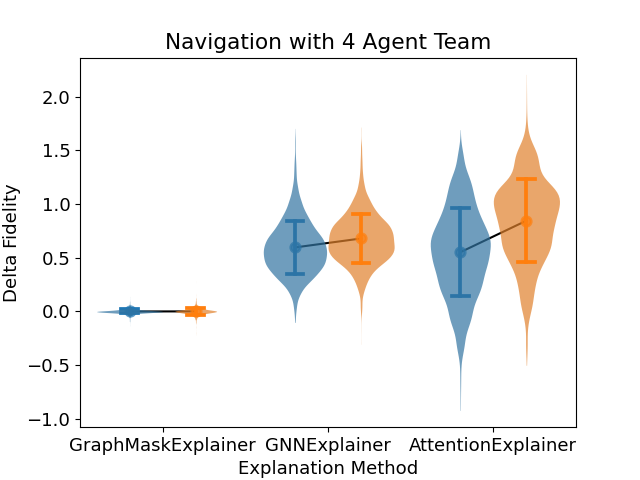}
    \includegraphics[trim={0.5cm 0.65cm 1.55cm 0.8cm},clip,width=0.317\linewidth]{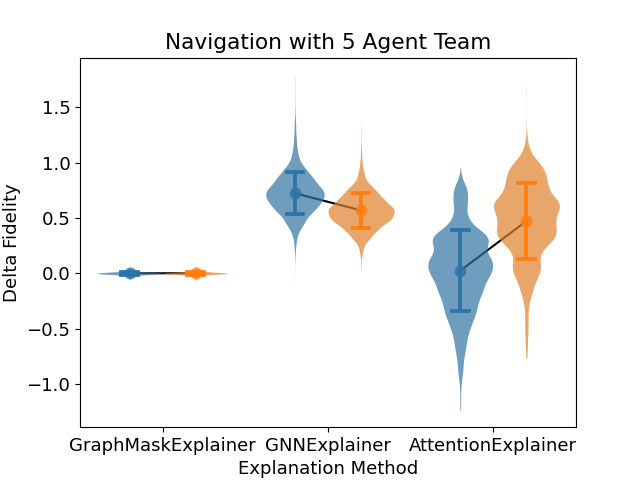}
    \includegraphics[trim={0.1cm 0.65cm 1.55cm 0.8cm},clip,width=0.325\linewidth]{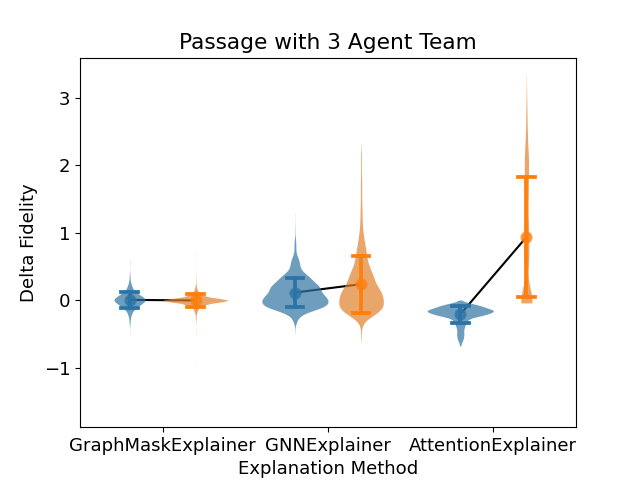}
    \includegraphics[trim={0.5cm 0.65cm 1.55cm 0.8cm},clip,width=0.317\linewidth]{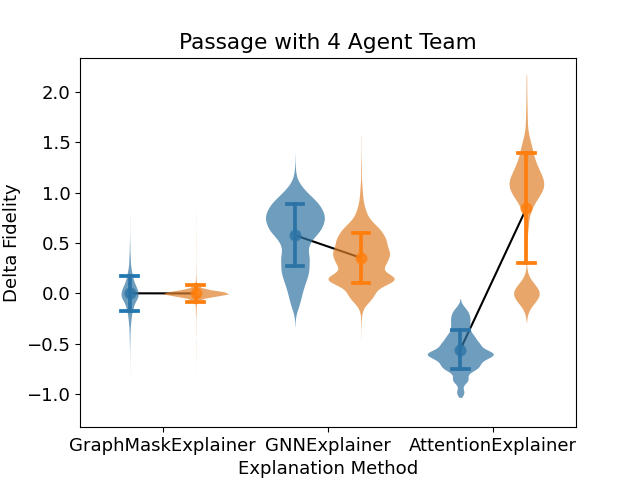}
    \includegraphics[trim={0.5cm 0.65cm 1.55cm 0.8cm},clip,width=0.317\linewidth]{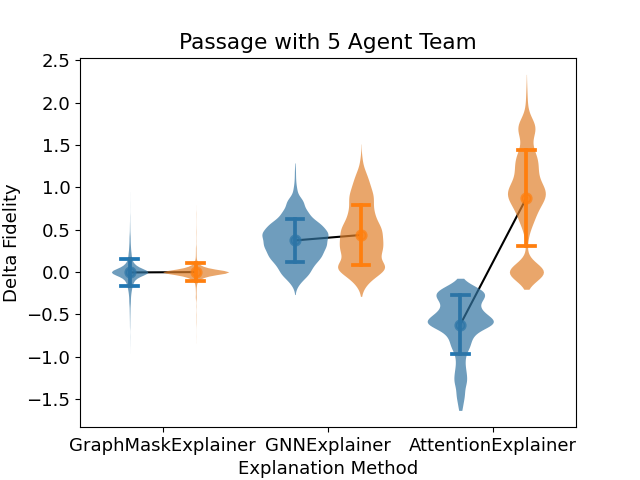}
    \includegraphics[trim={0.1cm 0.65cm 1.55cm 0.8cm},clip,width=0.325\linewidth]{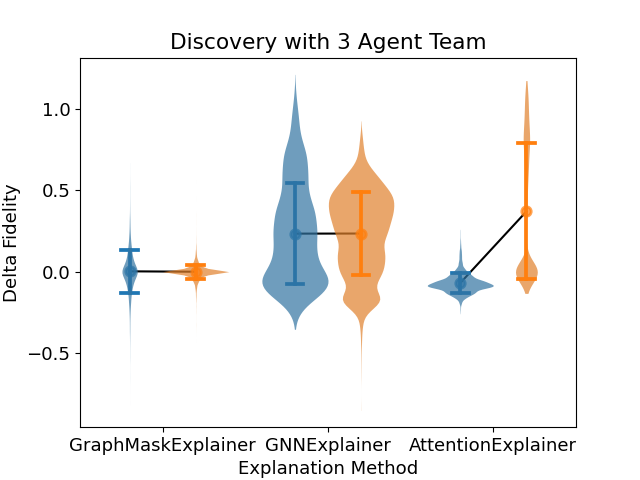}
    \includegraphics[trim={0.5cm 0.65cm 1.55cm 0.8cm},clip,width=0.317\linewidth]{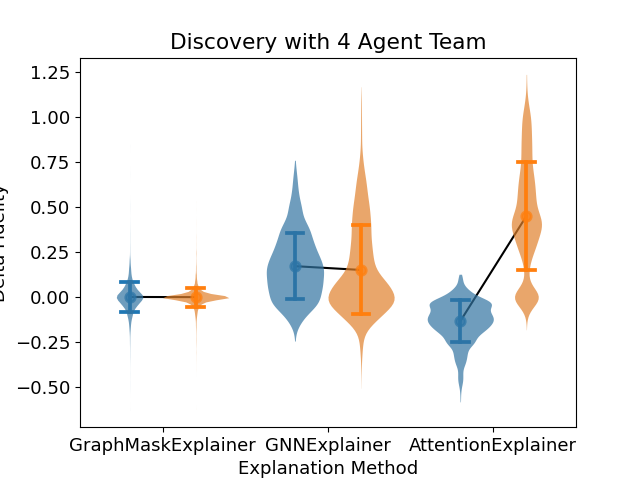}
    \includegraphics[trim={0.5cm 0.65cm 1.55cm 0.8cm},clip,width=0.317\linewidth]{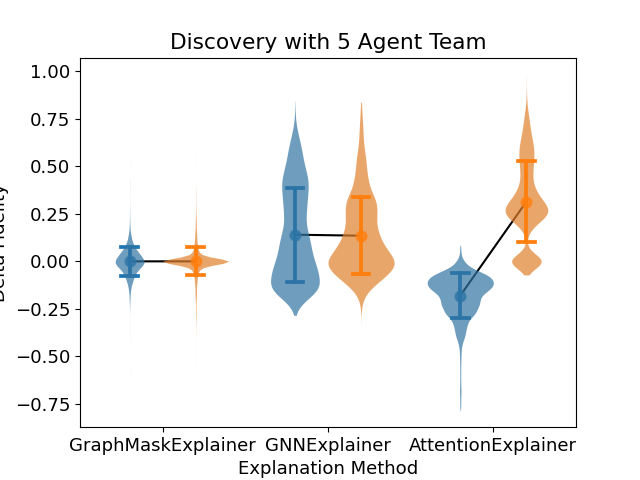}
    
    \caption{Delta Fidelity ($\uparrow$, \sectionautorefname~\ref{fidelityResults}) of explanations generated by three explainers \textcolor{blue}{\textbf{without (blue)}} and \textcolor{orange}{\textbf{with (orange)}} proposed regularization across three tasks (rows) and three team sizes (columns). Higher Delta Fidelity is better.
    }
    \label{fig:deltaFidelity}
\end{figure}

\subsection{Results on Explanation Fidelity}
\label{fidelityResults}
\figureautorefname~\ref{fig:deltaFidelity} shows the delta fidelity results for each of the three team sizes and each of the three tasks. Note that a higher delta fidelity is desirable, as this indicates that the explanation subgraph is both necessary and sufficient.

\textit{Comparing explainers}: We first analyze each explanation method without attention entropy with respect to delta fidelity (blue violin plots in \figureautorefname~\ref{fig:deltaFidelity}). We observe that GraphMask tends to have the least delta fidelity, followed by AttentionExplainer, with GNNExplainer yielding the highest delta fidelity. The low delta fidelity of GraphMask is likely due to its sole reliance on hard binary masks, which constrains the space of possible subgraph explanations. 
\sk{
GNNExplainer and AttentionExplainer likely outperform than GraphMask since they produce more expressive subgraphs and allow for varying degrees of inter-agent influences as they use soft edge masks (c.f. binary hard-masks in GraphMask). Unlike AttentionExplainer, GNNExplainer is compatible with any GNN (i.e. GCN, GAT, etc) and does not rely on the model being self-interpretable. Without attention entropy regularization, the attention values are more noisy and likely cause AttentionExplainer to perform worse than GNNExplainer. 
}
These observations suggest that GNNExplainer provides the best explanations with respect to delta fidelity when employed out-of-the-box.

\textit{Impact of regularization}: Now we examine the impact of attention entropy minimization on each explainer's fidelity measures and also compare them with one another. We observe that the regularization has had no discernible impact on GraphMask as it still tends to have the worst delta fidelity measures. This is because GraphMask gains an improvement in negative fidelity, but also incurs a deterioration in positive fidelity (see appendix). This means that the explanations generated by GraphMask on a model trained with attention entropy minimization contain a larger subset of the salient edges but do not capture all the salient edges due to the restriction of hard binary masks. GNNExplainer and AttentionExplainer continue to perform better than GraphMask after the regularization. 

However, unlike without attention entropy minimization, AttentionExplainer now outperforms GNNExplainer. 
This is likely because inclusion of attention entropy boosts the correlation between the attention values (which can be interpreted as a subgraph) and the GNN model behavior, in line with the insights from the theoretical analysis. 
Further, entropy minimization likely obtains mixed results for GNNExplainer with respect to fidelity due to the difficulty of the optimization problem \cite{gnnexplainer}. The original objective of GNNExplainer is intractable to solve, requiring assumptions that tend to work well for the graph datasets for which it was designed. 
Thus, the inclusion of attention entropy minimization may not inherently improve the optimization landscape that GNNExplainer attempts to solve despite the fact that the subgraphs induced by the attention values themselves better represent the underlying model. 

\begin{figure}[tb]
    \centering
    \includegraphics[trim={0.1cm 0.65cm 1.55cm 0.8cm},clip,width=0.325\linewidth]{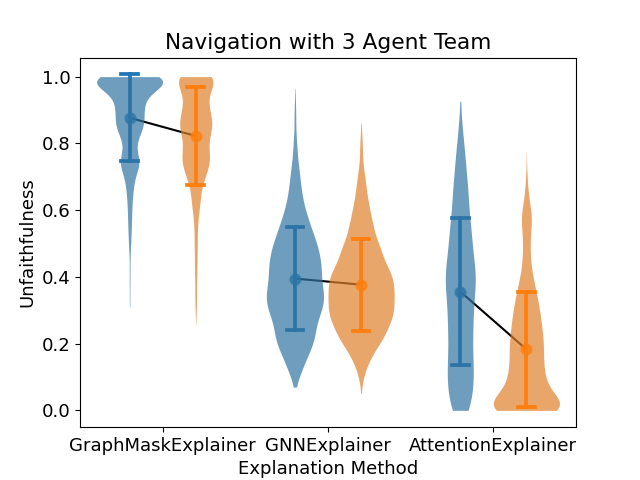}
    \includegraphics[trim={0.5cm 0.65cm 1.55cm 0.8cm},clip,width=0.317\linewidth]{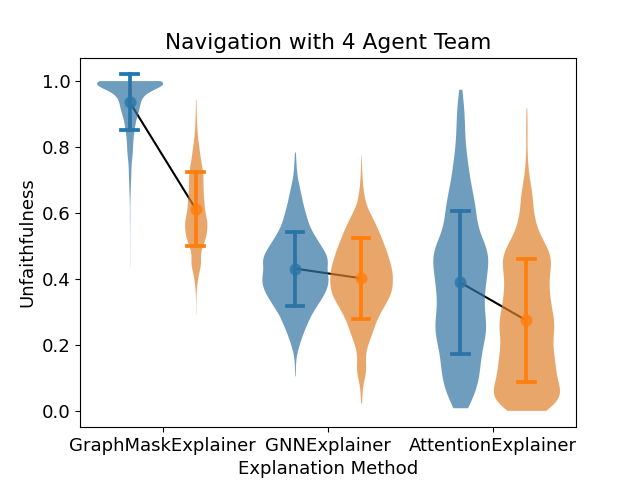}
    \includegraphics[trim={0.5cm 0.65cm 1.55cm 0.8cm},clip,width=0.317\linewidth]{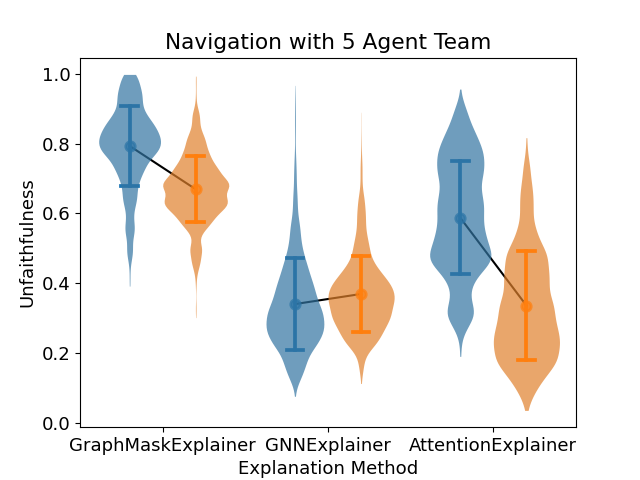}
    
    \includegraphics[trim={0.1cm 0.65cm 1.55cm 0.8cm},clip,width=0.325\linewidth]{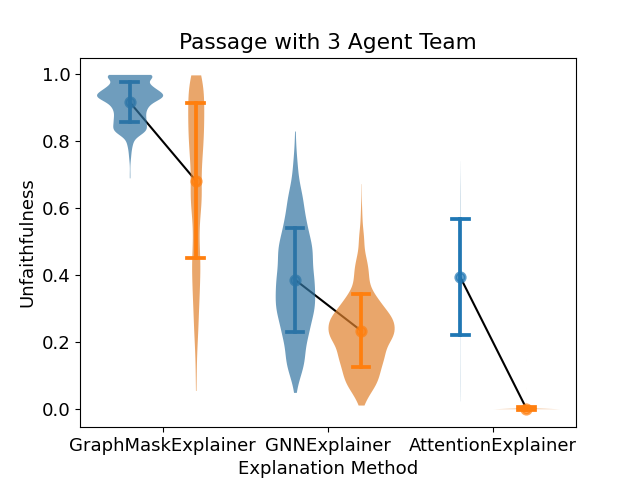}
    \includegraphics[trim={0.5cm 0.65cm 1.55cm 0.8cm},clip,width=0.317\linewidth]{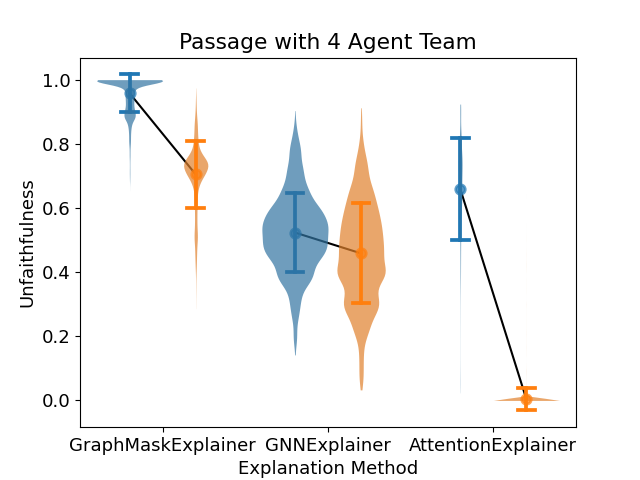}
    \includegraphics[trim={0.5cm 0.65cm 1.55cm 0.8cm},clip,width=0.317\linewidth]{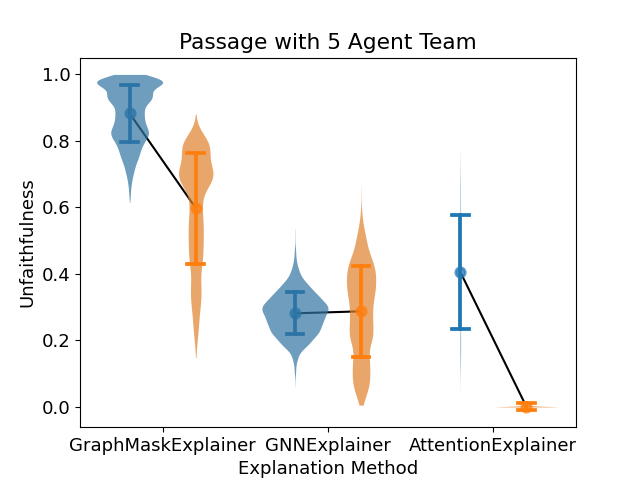}
    
    \includegraphics[trim={0.1cm 0.65cm 1.55cm 0.8cm},clip,width=0.325\linewidth]{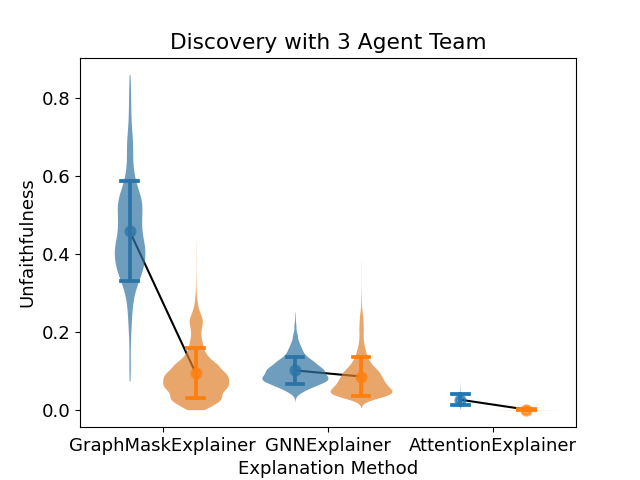}
    \includegraphics[trim={0.5cm 0.65cm 1.55cm 0.8cm},clip,width=0.317\linewidth]{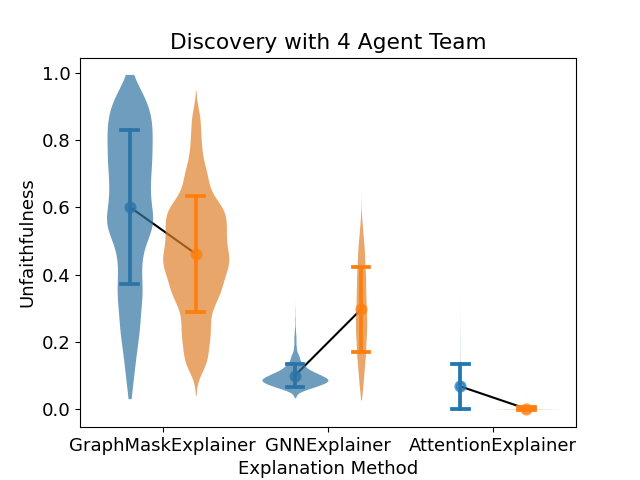}
    \includegraphics[trim={0.5cm 0.65cm 1.55cm 0.8cm},clip,width=0.317\linewidth]{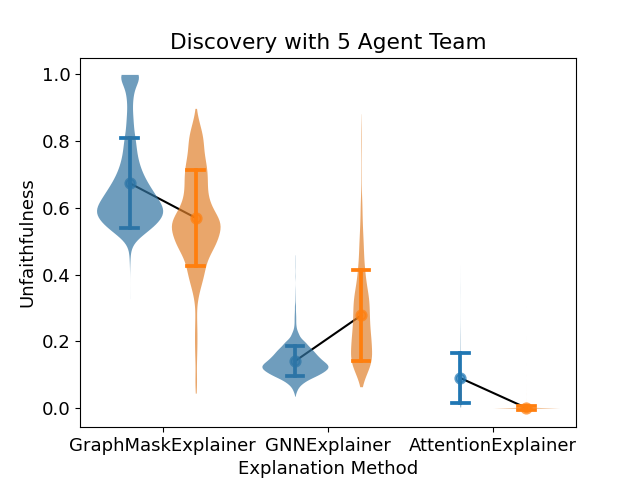}
    \caption{Unfaithfulness ($\downarrow$, \sectionautorefname~\ref{unfaithResults}) of explanations generated by three explainers \textcolor{blue}{\textbf{without (blue)}} and \textcolor{orange}{\textbf{with (orange)}} proposed regularization across three tasks (rows) and three team sizes (columns). Lower unfaithfulness is better.
    }
    \label{fig:unfaithfulness}
\end{figure}

\begin{table*}[bt]
\scriptsize
\centering
\begin{tabular}{|c|c|c|c|c|c|c|}
\hline
\multicolumn{7}{|c|}{\textbf{Blind Navigation}} \\
\hline
\multirow{2}{*}{Performance Metric} & \multicolumn{2}{c|}{N = 3} & \multicolumn{2}{c|}{N = 4} & \multicolumn{2}{c|}{N = 5} \\
    \cline{2-7}
     &  No Att. Ent. Min. & Att. Ent. Min. & No Att. Ent. Min. & Att. Ent. Min. & No Att. Ent. Min. & Att. Ent. Min. \\
\hline
Reward ($\uparrow$) & 2.32 $\pm$ 0.64 & 2.29 $\pm$ 0.58 & 3.05 $\pm$ 0.82 & 3.03 $\pm$ 0.70 & 4.75 $\pm$ 1.10 & 4.64 $\pm$ 2.04 \\
\hline
Success Rate ($\uparrow$) & 0.99 $\pm$ 0.10 & 0.99 $\pm$ 0.10 & 0.96 $\pm$ 0.21 & 0.94 $\pm$ 0.24 & 0.80 $\pm$ 0.40 & 0.90 $\pm$ 0.30 \\
\hline
No Agent Coll. Rate ($\uparrow$) & 1.0 $\pm$ 0.0 & 1.0 $\pm$ 0.07 & 0.98 $\pm$ 0.14 & 1.0 $\pm$ 0.07 & 0.94 $\pm$ 0.22 & 0.94 $\pm$ 0.24 \\
\hline
Makespan ($\downarrow$) & 48.45 $\pm$ 24.90 & 48.99 $\pm$ 23.47 & 61.41 $\pm$ 27.47 & 64.18 $\pm$ 30.69 & 200.92 $\pm$ 116.44 & 175.86 $\pm$ 104.04 \\
\hline
\hline
\multicolumn{7}{|c|}{\textbf{Passage}} \\
\hline
\multirow{2}{*}{Performance Metric} & \multicolumn{2}{c|}{N = 3} & \multicolumn{2}{c|}{N = 4} & \multicolumn{2}{c|}{N = 5} \\
    \cline{2-7}
     &  No Att. Ent. Min. & Att. Ent. Min. & No Att. Ent. Min. & Att. Ent. Min. & No Att. Ent. Min. & Att. Ent. Min. \\
\hline
Reward ($\uparrow$) & 54.5 $\pm$ 12.7 & 55.6 $\pm$ 23.1 & 44.5 $\pm$ 17.3 & 32.1 $\pm$ 27.1 & 37.9 $\pm$ 24.3 & 33.2 $\pm$ 21.7 \\
\hline
Success Rate ($\uparrow$) & 1.0 $\pm$ 0.0 & 1.0 $\pm$ 0.0 & 0.69 $\pm$ 0.3 & 1.0 $\pm$ 0.02 & 0.91 $\pm$ 0.24 & 0.99 $\pm$ 0.08 \\
\hline
No Agent Coll. Rate ($\uparrow$) & 0.30 $\pm$ 0.50 & 0.69 $\pm$ 0.50 & 0.58 $\pm$ 0.50 & 0.32 $\pm$ 0.50 & 0.11 $\pm$ 0.31 & 0.22 $\pm$ 0.41 \\
\hline
No Object Coll. Rate ($\uparrow$) & 0.0 $\pm$ 0.0 & 0.06 $\pm$ 0.2 & 0.0 $\pm$ 0.0 & 0.0 $\pm$ 0.0 & 0.0 $\pm$ 0.0 & 0.0 $\pm$ 0.0 \\
\hline
Makespan ($\downarrow$) & 45.445 $\pm$ 21.8 & 49.015 $\pm$ 14.4 & 119.505 $\pm$ 42.4 & 119.675 $\pm$ 35.3 & 84.885 $\pm$ 37.7 & 100.2 $\pm$ 33.26 \\
\hline
\hline
\multicolumn{7}{|c|}{\textbf{Discovery}} \\
\hline
\multirow{2}{*}{Performance Metric} & \multicolumn{2}{c|}{N = 3} & \multicolumn{2}{c|}{N = 4} & \multicolumn{2}{c|}{N = 5} \\
    \cline{2-7}
     &  No Att. Ent. Min. & Att. Ent. Min. & No Att. Ent. Min. & Att. Ent. Min. & No Att. Ent. Min. & Att. Ent. Min. \\
\hline
Reward ($\uparrow$) & 0.92 $\pm$ 0.27 & 0.88 $\pm$ 0.35 & 0.99 $\pm$ 0.12 & 0.96 $\pm$ 0.23 & 0.92 $\pm$ 0.29 & 0.95 $\pm$ 0.23 \\
\hline
Success Rate ($\uparrow$) & 0.92 $\pm$ 0.27 & 0.87 $\pm$ 0.34 & 0.99 $\pm$ 0.12 & 0.95 $\pm$ 0.22 & 0.92 $\pm$ 0.28 & 0.95 $\pm$ 0.23 \\
\hline
Makespan ($\downarrow$) & 112.38 $\pm$ 135.85 & 145.19 $\pm$ 150.28 & 95.16 $\pm$ 103.32 & 134.79 $\pm$ 120.83 & 113.25 $\pm$ 133.33 & 108.95 $\pm$ 116.55 \\
\hline
\end{tabular}
\caption{Attention entropy minimization results in little to no degradation in task performance across all tasks and team sizes.}
\label{table:AllTasksPerformance}
\end{table*}

\subsection{Results on Explanation Faithfulness}
\label{unfaithResults}

\figureautorefname~\ref{fig:unfaithfulness} shows the unfaithfulness results for each of the three team sizes and each of the three tasks. Note that a \textit{lower} unfaithfulness is generally desirable, as this indicates that the explanation subgraph is more faithful to the underlying prediction distribution.

\textit{Comparing explainers}: We first analyze each explanation method without attention entropy with respect to unfaithfulness (blue violin plots in \figureautorefname~\ref{fig:unfaithfulness}). We observe that, out of the three explanation methods, GraphMask tends to have the worst unfaithfulness measures. Of the 9 experimental configurations, 5 configurations show AttentionExplainer having marginally better unfaithfulness measures compared to GNNExplainer, and the remaining 4 configurations show the opposite. Once again, the high unfaithfulness measures from GraphMask are likely due to its use of hard binary masks. 
\sk{
Especially when the attention values are diffuse, GNNExplainer and AttentionExplainer are more likely to be able to capture varying degrees of inter-agent influences via soft edges. On the other hand, GraphMask either preserves or drops an edge, making it difficult to capture more diffuse inter-agent influences, which are more likely to occur when the model is trained without regularization. 
}

\textit{Impact of regularization}: Now we examine the impact of attention entropy minimization on each explainer's faithfulness and also compare them with one another. The regularization consistently reduces GraphMask's unfaithfulness. This is likely due to the regularization making the attention distribution sparser, resulting in agent-agent influences that are closer to a subgraph composed of only a binary adjacency matrix. As such, GraphMask can capture more of the salient edges even when employing a hard mask. However, this boost from regularization seems to be insufficient to overcome all the limitations of hard masks since GraphMask continues to underperform the other two explainers with respect to faithfulness. In contrast, the regularization significantly improves the faithfulness of AttentionExplainer across all team sizes and tasks, enabling it to now consistently outperform GNNExplainer. Once again, this can likely be attributed to the fact that the minimizing attention entropy boosts the correlation between the subgraph induced by the attention values and the inter-agent influences considered by the underlying model. Similar to what we observed for fidelity, minimizing attention entropy seems to have a mixed impact on GNNExplainer's faithfulness.


\subsection{Impact on Task Performance}
We also measured the task performance comparison between the presence and absence of attention entropy minimization when training the policy (see Appendix for metrics). Generally, it is common to expect some performance loss in pursuit of explainability, but a model that is more explainable but does not perform the task well is not useful. Moreover, a potential concern of augmenting the RL objective with attention entropy is the reduction in policy performance. Reassuringly, \tableautorefname~\ref{table:AllTasksPerformance} shows that, in general, little to no task performance is lost when including attention entropy minimization into the learning objective.

\subsection{Additional Experiments}
We conduct two additional experimental studies in conjunction with the above evaluation. The first study looks at the explanation quality over the course of model training, results can be found in \ref{subsec:explanation-quality-training}. The second study assesses the explanation quality in zero-shot deployment of trained policies on larger team sizes than the training team size, results can be found in \ref{subsec:explanation-quality-zero-shot}. The trends and findings from these studies are generally similar to the trends and findings of the aforementioned results. In particular, we continue to observe that GraphMask improves in faithfulness and AttentionExplainer improve in overall explanation quality while GNNExplainer garners mixed improvements when the model is trained using attention entropy minimization.


\section{Conclusion}
Our work lays the foundation for the adoption of GNN explainers to explain graph-based multi-agent policies. We conducted the first empirical evaluation of GNN explanation methods in explaining policies trained on established multi-agent tasks across varying team sizes. We find that existing GNN explanation methods have the potential to identify the most influential communication channels impacting the team's decisions. Further, we proposed attention entropy minimization as an effective regularization to improve explanation quality for graph attention-based policies, with theoretical motivations grounded in maximizing the separation between the subgraph and its complement. Future work could improve GNN explanation methods for MARL settings and the interpretability of subgraph explanations with natural language summarization.

\section*{Impact Statement}


This paper presents work whose goal is to advance the field of 
Machine Learning. In particular, the goal is to advance the multi-agent and multi-robot communities by providing an evaluation and improvement of a desirable explanation paradigm for multi-agent coordination.
There are many potential societal consequences 
of our work, none which we feel must be specifically highlighted here.


\balance
\bibliography{example_paper}
\bibliographystyle{icml2025}

\newpage
\appendix
\onecolumn
\section{Appendix}
\subsection{Training Details}
\sk{
In alignment with many MARL-based works \cite{nayak2023scalable, blumenkamp2021emergence, passageProrok, yang2023learning, he2023multi, escudie2024attention}, we adopt a centralized training decentralized execution (CTDE) paradigm to train policies. Specifically, we use multi-agent proximal policy optimization (MAPPO) \cite{yu2022surprising} with a centralized critic and homogeneous agents, which has been used in \cite{nayak2023scalable, yang2023learning, he2023multi, escudie2024attention}. The critic is parameterized as a 2-layer fully-connected multilayer perceptron (MLP) with 32 units in each layer for the blind navigation task and the passage task, which we borrow from \cite{passageProrok}, and 128 units in each layer for the discovery task, which we found worked well across all team sizes without attention entropy regularization (see \sectionautorefname~\ref{subsec:entropy_reg}). Following \cite{yu2022surprising}, we incorporate generalized advantage estimation \cite{Schulmanetal_ICLR2016} and parameter sharing for the agents (though the policies are decentralized). We optimize the MAPPO loss term using ADAM optimizer \cite{kingma2014adam}, and implement the MARL training and evaluation using PyTorch \cite{paszke2019pytorch} and PyTorch Geometric \cite{Fey/Lenssen/2019}.

\textbf{To ensure an unbiased evaluation of entropy regularization and different explanation methods, 
all our design choices for training (actor, critic, hyperparameters, etc.) are motivated by prior works or empirical performance of the policy without attention entropy minimization (from \sectionautorefname~\ref{subsec:entropy_reg}). In addition, we do not inform these design choices based on explanation quality either.
}
}

\subsection{Metrics for Task Performance}
We consider the following task performance metrics.

\textbf{Reward:} This is the average reward accrued across all agents, computed as $\frac{1}{N}\sum^N_{n=1}\sum^T_{t=0} r_t(n)$ where $n \in N$ is agent index and $t \in T$ is the time index within an episode. 


\textbf{Success Rate:} For the blind navigation and passage task, the success rate of an episode is defined by the fraction of the total number of agents who reach their assigned goals. 
An episode of the discovery task is considered successful if the landmark is discovered before the episode terminates.

\textbf{No Agent Collision Fraction:} Across all tasks, this metric is defined as the fraction of episodes that result in no agent-agent collisions, averaged across random initial states. 

\textbf{No Object Collision Fraction:} This is the fraction of episodes in which no agent collides with an obstacle, averaged across multiple random initial states. 
This metric is only used in the passage task, since 
there are no obstacles in blind navigation and discovery tasks.

\textbf{Makespan:} This is the number of environment steps it takes for the team to complete the task, with a maximum value defined by episode length.
We consider the blind navigation and passage tasks complete when the last agent reaches its goal location. 
We consider the discovery task complete when the landmark is discovered.

\subsection{Navigation Environment Training Details}
We train all policies for the navigation task with the hyperparameters in Table \ref{table:navigation-hyperparameters}. We set the weight of the attention entropy term to 10.
\begin{table}[h!]
\centering
\begin{tabular}{|c|c|}
\hline
\textbf{Hyperparameter} & \textbf{Value} \\ \hline
Learning rate & 0.0003 \\ \hline
Gamma & 0.99999 \\ \hline
Lambda & 0.9 \\ \hline
Entropy epsilon & 0.0001 \\ \hline
Clip epsilon & 0.2 \\ \hline
Critic loss type & Smooth L1 \\ \hline
Normalize advantage & False \\ \hline
Number of epochs & 30 \\ \hline
Max gradient norm & 1.0 \\ \hline
Minibatch size & 800 \\ \hline
Collector iterations & 150 \\ \hline
Frames per batch & 18000 \\ \hline
\end{tabular}
\caption{Navigation Hyperparameters}
\label{table:navigation-hyperparameters}
\end{table}

\subsection{Passage Environment Training Details}
We train all policies for the passage task with the hyperparameters in Table \ref{table:passage-hyperparameters}. We set the weight of the attention entropy term to 50.
\begin{table}[h!]
\centering
\begin{tabular}{|c|c|}
\hline
\textbf{Hyperparameter} & \textbf{Value} \\ \hline
Learning rate & 0.00005 \\ \hline
Gamma & 0.99999 \\ \hline
Lambda & 0.9 \\ \hline
Entropy epsilon & 0.0001 \\ \hline
Clip epsilon & 0.2 \\ \hline
Critic loss type & Smooth L1 \\ \hline
Normalize advantage & False \\ \hline
Number of epochs & 30 \\ \hline
Max gradient norm & 1.0 \\ \hline
Minibatch size & 800 \\ \hline
Collector iterations & 100 \\ \hline
Frames per batch & 60000 \\ \hline
\end{tabular}
\caption{Passage Hyperparameters}
\label{table:passage-hyperparameters}
\end{table}


\subsection{Discovery Environment Training Details}
We train all policies for the discovery task with the hyperparameters in Table \ref{table:discovery-hyperparameters}. We set the weight of the attention entropy term to 50.
\begin{table}[h!]
\centering
\begin{tabular}{|c|c|}
\hline
\textbf{Hyperparameter} & \textbf{Value} \\ \hline
Learning rate & 0.0007 \\ \hline
Gamma & 0.9999 \\ \hline
Lambda & 0.95 \\ \hline
Entropy epsilon & 0.0001 \\ \hline
Clip epsilon & 0.05 \\ \hline
Critic loss type & Smooth L1 \\ \hline
Normalize advantage & True \\ \hline
Number of epochs & 5 \\ \hline
Max gradient norm & 10 \\ \hline
Minibatch size & 10000 \\ \hline
Collector iterations & 1000 \\ \hline
Frames per batch & 10000 \\ \hline
\end{tabular}
\caption{Discovery Hyperparameters}
\label{table:discovery-hyperparameters}
\end{table}

\subsection{Positive Fidelity and Negative Fidelity}
\begin{figure}[!h]
    \centering
    \includegraphics[width=0.33\linewidth]{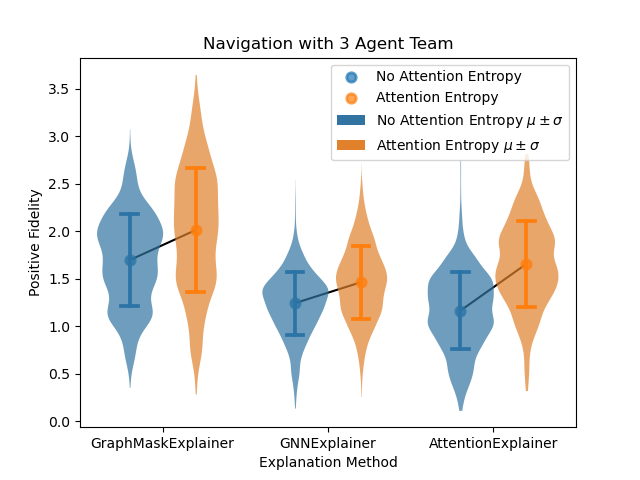}
    \includegraphics[width=0.33\linewidth]{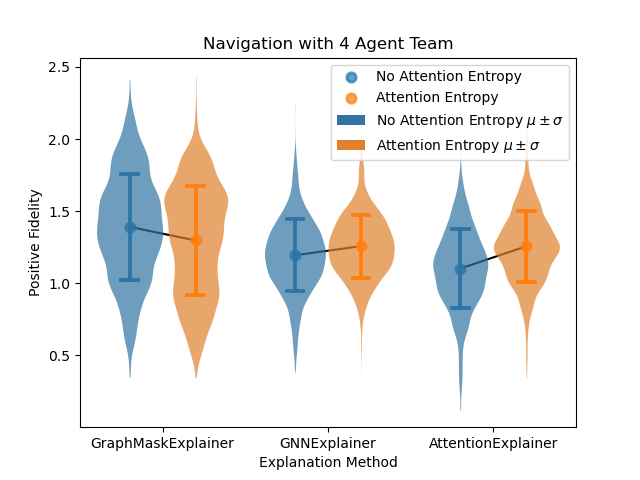}
    \includegraphics[width=0.33\linewidth]{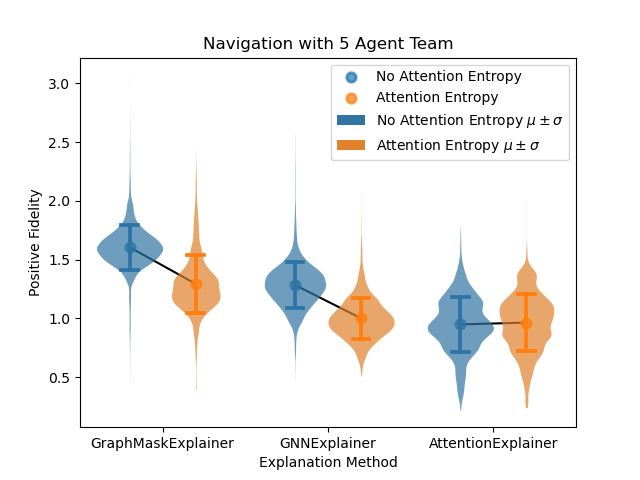}
    \includegraphics[width=0.33\linewidth]{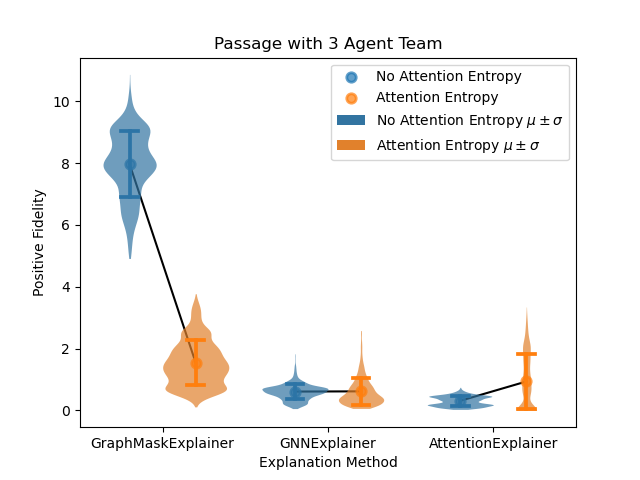}
    \includegraphics[width=0.33\linewidth]{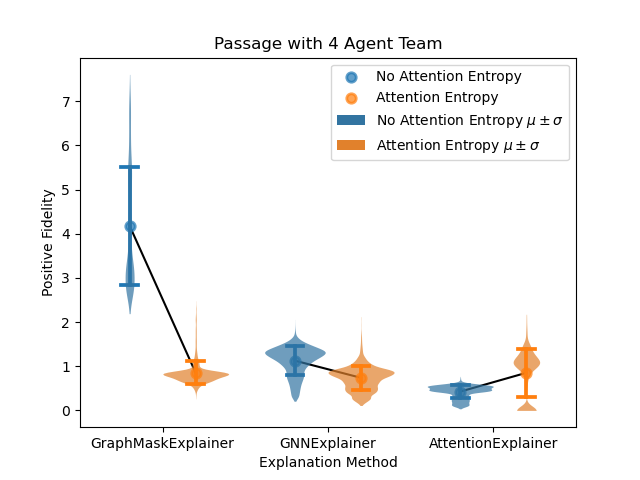}
    \includegraphics[width=0.33\linewidth]{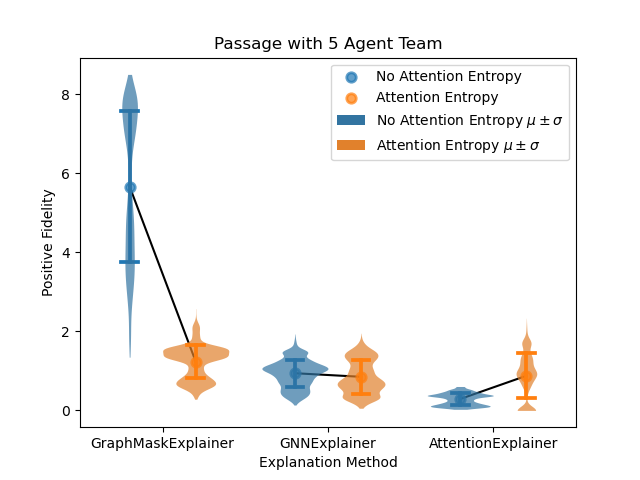}
    \includegraphics[width=0.33\linewidth]{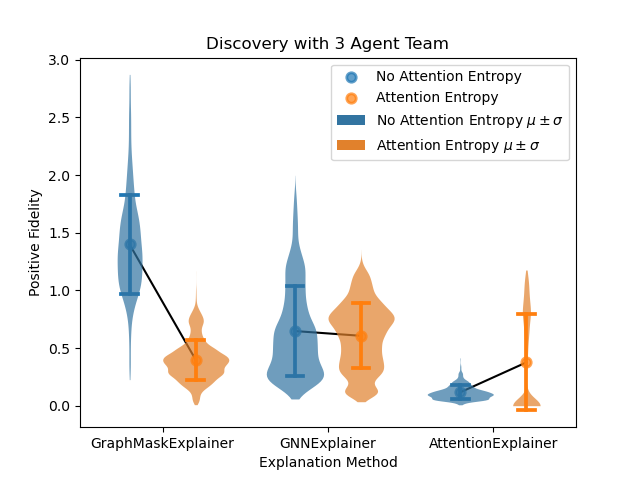}
    \includegraphics[width=0.33\linewidth]{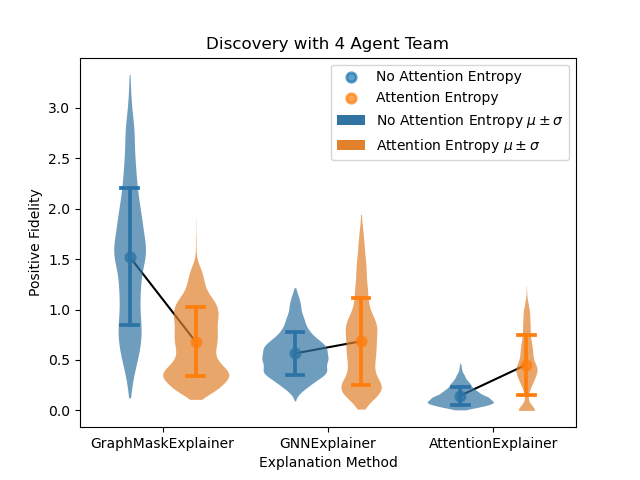}
    \includegraphics[width=0.33\linewidth]{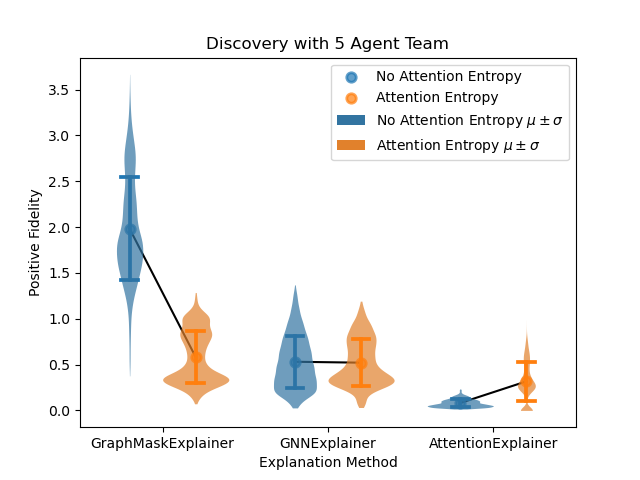}
    \caption{Positive Fidelity ($\uparrow$) of explanations generated by three explanation methods with and without proposed regularization, across three team sizes. \textit{Top row}: Blind navigation task, \textit{Middle row}: Passage task, \textit{Bottom row}: Discovery task. Higher is better.}
    \label{fig:positiveFidelity}
\end{figure}

\begin{figure}[!h]
    \centering
    \includegraphics[width=0.33\linewidth]{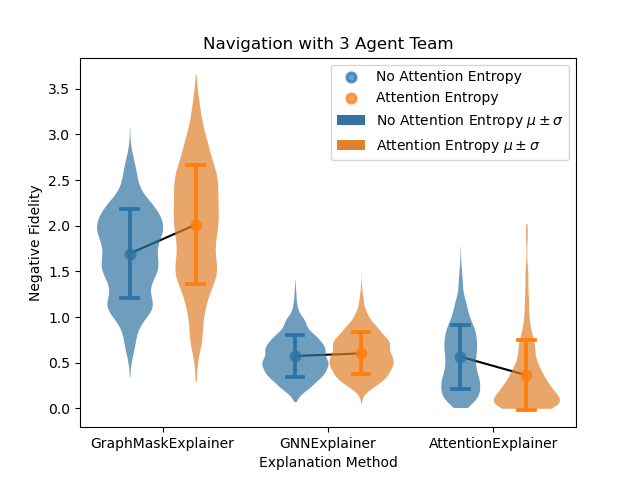}
    \includegraphics[width=0.33\linewidth]{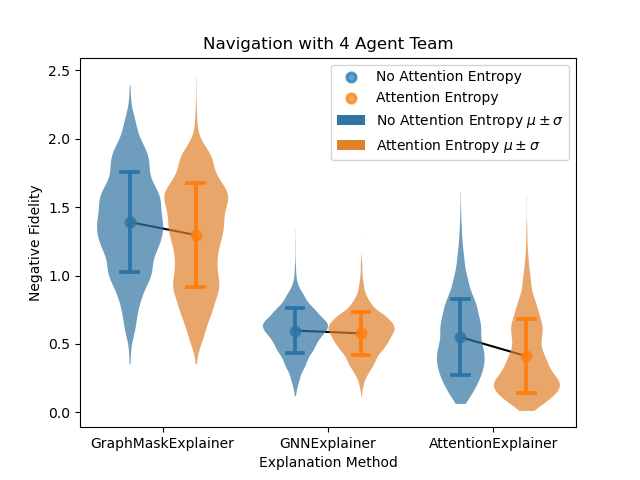}
    \includegraphics[width=0.33\linewidth]{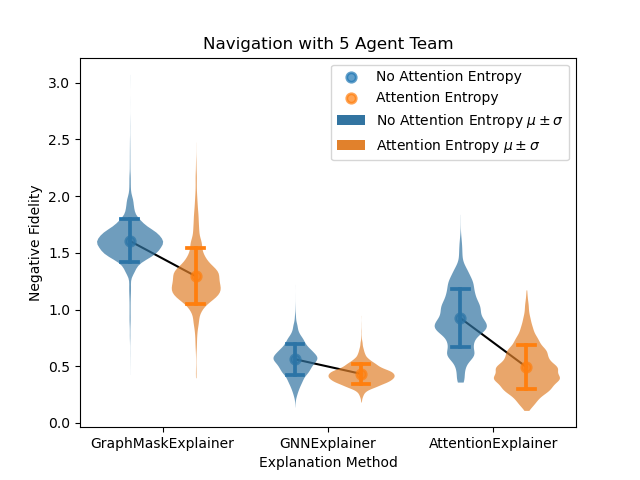}
    \includegraphics[width=0.33\linewidth]{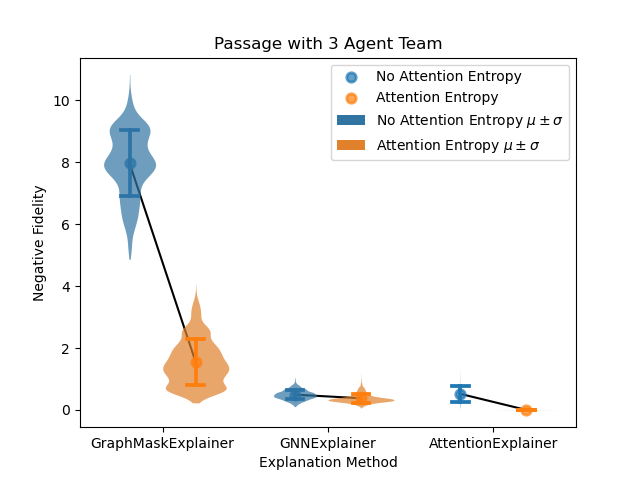}
    \includegraphics[width=0.33\linewidth]{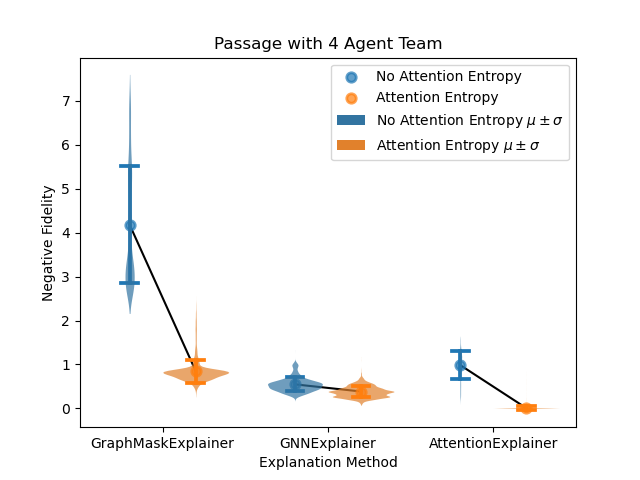}
    \includegraphics[width=0.33\linewidth]{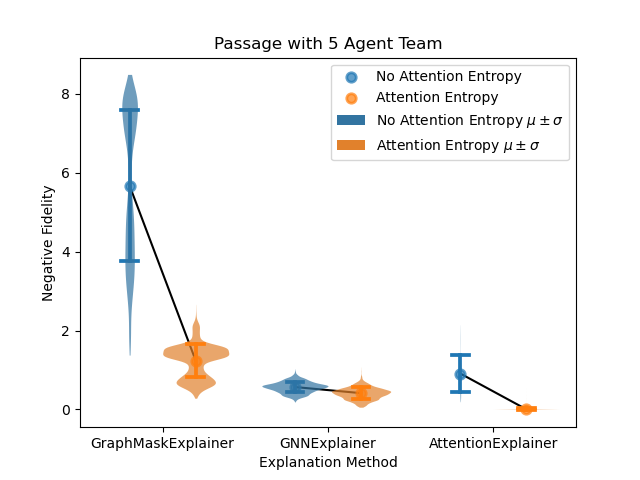}
    \includegraphics[width=0.33\linewidth]{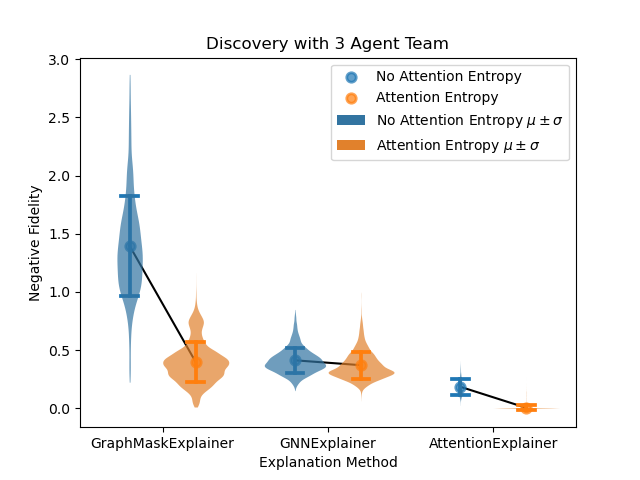}
    \includegraphics[width=0.33\linewidth]{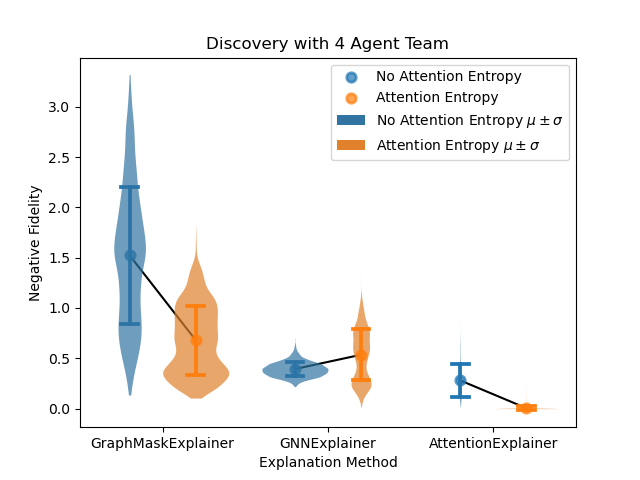}
    \includegraphics[width=0.33\linewidth]{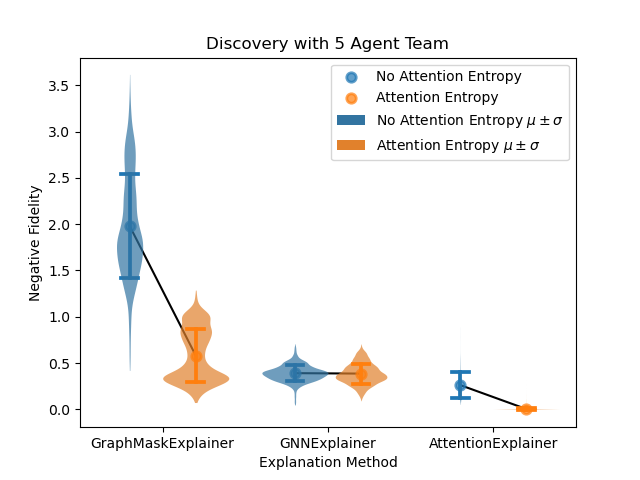}
    \caption{Negative Fidelity ($\downarrow$) of explanations generated by three explanation methods with and without proposed regularization, across three team sizes. \textit{Top row}: Blind navigation task, \textit{Middle row}: Passage task, \textit{Bottom row}: Discovery task. Lower is better.}
    \label{fig:negativeFidelity}
\end{figure}

\figureautorefname~\ref{fig:positiveFidelity} shows the positive fidelity measures and \figureautorefname~\ref{fig:negativeFidelity} shows the negative fidelity measures. When the model is trained with attention entropy minimization, GraphMask tends to improve in negative fidelity, but regresses in positive fidelity, yielding little net benefit in delta fidelity. GNNExplainer has mixed improvements for both positive and negative fidelity, which yields mixed results in delta fidelity. AttentionExplainer consistently improves in positive and negative fidelity, which accounts for the large improvement in delta fidelity that is observed.

\subsection{Explanation Quality Over Training Iterations}
\label{subsec:explanation-quality-training}
\begin{figure}[!h]
    \centering
    \includegraphics[width=0.33\linewidth]{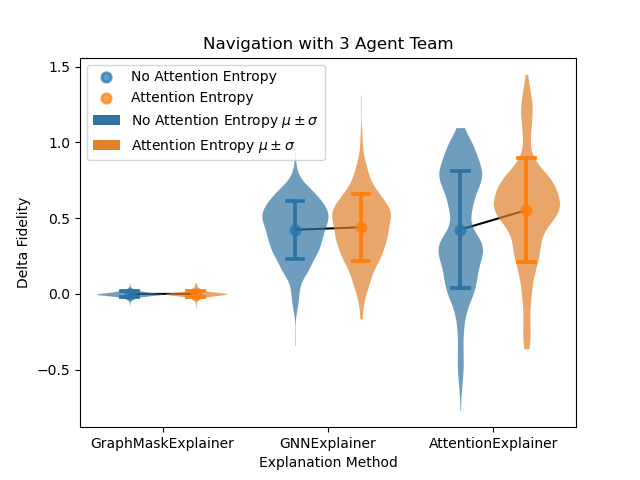}
    \includegraphics[width=0.33\linewidth]{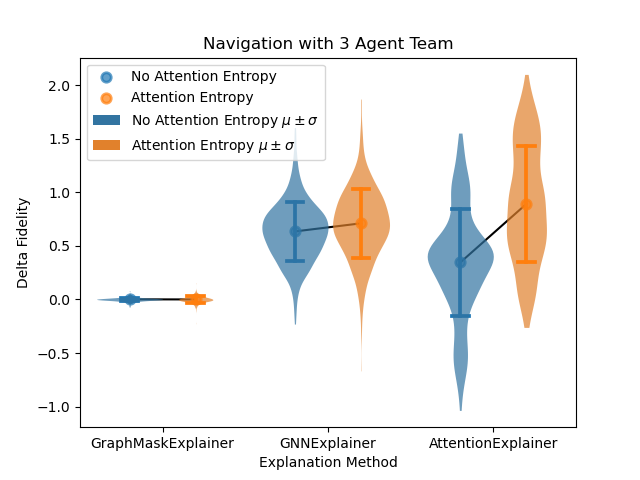}
    \includegraphics[width=0.33\linewidth]{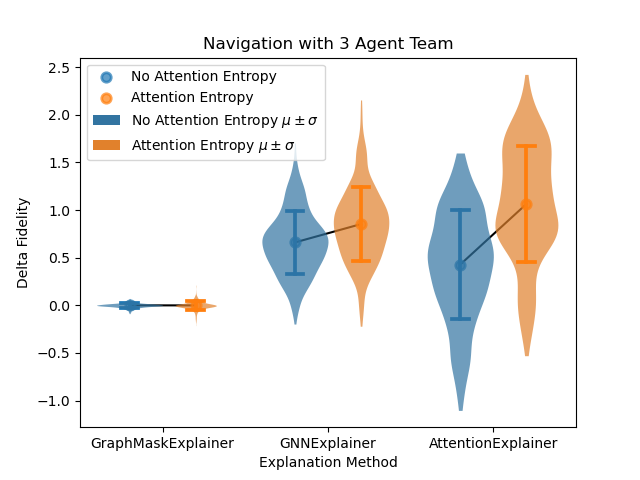}
    \includegraphics[width=0.33\linewidth]{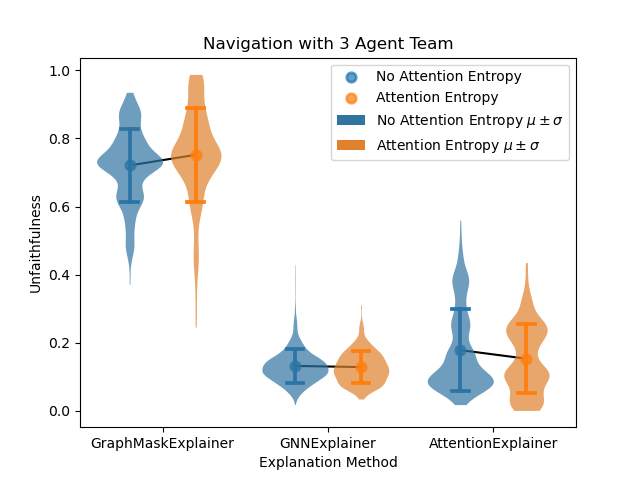}
    \includegraphics[width=0.33\linewidth]{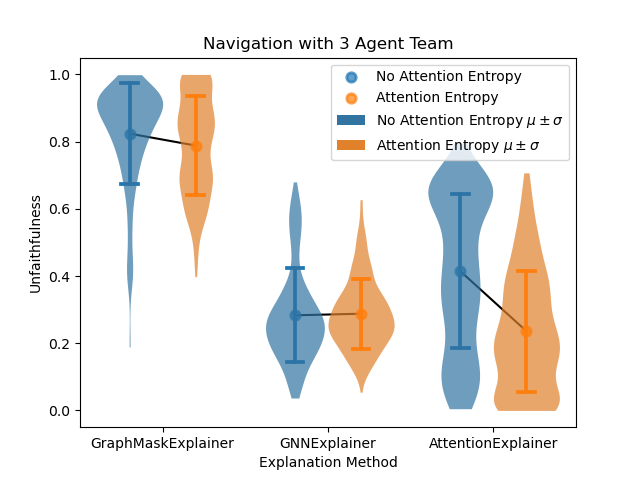}
    \includegraphics[width=0.33\linewidth]{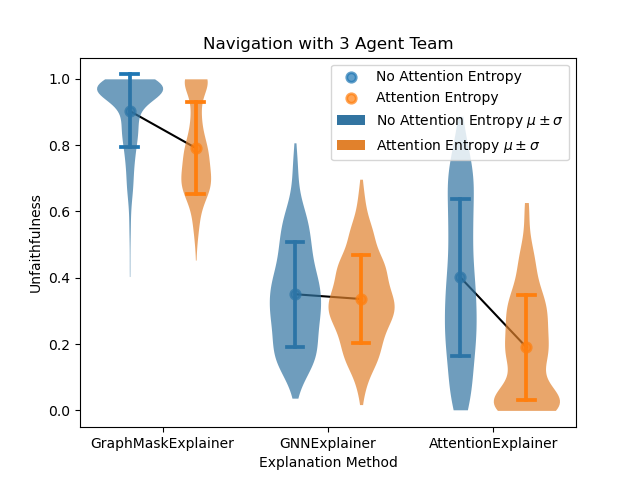}
    \caption{Delta Fidelity (top, $\uparrow$) and Unfaithfulness (bottom, $\downarrow$) of explanations generated by three explanation methods with and without proposed regularization at 20\% (left), 50\% (middle), and 80\% (right) training completion.}
    \label{fig:explanationQualityOverTraining}
\end{figure}

We analyze the explanation quality across training iterations to evaluate what post-hoc explainers are best suited for mediocre performing policies or for debugging models that are still training. \figureautorefname~\ref{fig:explanationQualityOverTraining} shows the delta fidelity and unfaithfulness measures for 10\%, 50\%, and 90\% training for a 3 agent GNN-based policy for the blind navigation task. This provides some insight into the evolution of the explanation quality as task reward optimization and attention entropy minimization occurs.

\subsubsection{Delta Fidelity}
As seen before, GraphMask does not gain any improvement in Delta Fidelity due to the sole reliance on hard binary masks and the approximately equal trade-off between Positive and Negative Fidelity that occurs when GraphMask is able to capture a larger subset of important agent-agent influences but still misses the other remaining important agent-agent influences. GNNExplainer and AttentionExplainer both improve over the course of training. Moreover, both GNNExplainer and AttentionExplainer benefit from attention entropy minimization, with AttentionExplainer inhereting a larger improvement due to attention entropy minimization. Finally, while GraphMask provides the least fidelity explanation regardless of the presence or absence of regularization, GNNExplainer performs the best across models without attention entropy minimization while AttentionExplainer performs the best across models with attention entropy minimization. AttentionExplainer, evaluated on a model trained with attention entropy minimization, yields the highest fidelity explanations overall.

\subsubsection{Unfaithfulness}
As seen before, GraphMask improves in faithfulness when the model is trained with attention entropy minimization, which is likely due to attention entropy minimization making the distribution sparser, resulting in agent-agent influences that are closer to a subgraph composed of only a binary adjacency matrix. Still, GraphMask does not achieve as faithful of explanations as GNNExplainer and AttentionExplainer. This is once again attributed to the limited representation capacity of GraphMask subgraphs since they are constrained to be hard binary masks, and attention entropy minimization can only ameliorate this, to an extent. In addition, we observe that the faithfulness of AttentionExplainer improves due to attention entropy minimization, while GNNExplainer has negligible to slight improvements in faithfulness due to attention entropy minimization. Finally, while GraphMask provides the least faithful explanations overall regardless of the presence or absence of regularization, GNNExplainer performs the best across models without attention entropy minimization while AttentionExplainer performs the best across models with attention entropy minimization. AttentionExplainer, evaluated on a model trained with attention entropy minimization, yields the most faithful (least unfaithful) explanations overall.

\subsubsection{Task Performance}
\tableautorefname~\ref{table:TrainingIterationPerformance} presents the task performance metrics for each of the aforementioned model checkpoints. Overall, the performance of the models trained with attention entropy minimization ranges from being comparable to outperforming the models trained without attention entropy minimization.

\begin{table*}[bt]
\scriptsize
\centering
\begin{tabular}{|c|c|c|c|c|c|c|}
\hline
\multicolumn{7}{|c|}{\textbf{Blind Navigation}} \\
\hline
\multirow{2}{*}{Performance Metric} & \multicolumn{2}{c|}{20\% Training Completed} & \multicolumn{2}{c|}{50\% Training Completed} & \multicolumn{2}{c|}{80\% Training Completed} \\
    \cline{2-7}
     &  No Att. Ent. Min. & Att. Ent. Min. & No Att. Ent. Min. & Att. Ent. Min. & No Att. Ent. Min. & Att. Ent. Min. \\
\hline
Reward ($\uparrow$) & 2.34 $\pm$ 0.69 & 2.11 $\pm$ 0.77 & 2.37 $\pm$ 0.65 & 2.37 $\pm$ 0.60 & 2.20 $\pm$ 0.81 & 2.39 $\pm$ 0.53 \\
\hline
Success Rate ($\uparrow$) & 0.22 $\pm$ 0.41 & 0.75 $\pm$ 0.43 & 0.97 $\pm$ 0.15 & 0.95 $\pm$ 0.21 & 0.97 $\pm$ 0.15 & 0.97 $\pm$ 0.15 \\
\hline
No Agent Coll. Rate ($\uparrow$) & 1.0 $\pm$ 0.0 & 1.0 $\pm$ 0.0 & 1.0 $\pm$ 0.0 & 1.0 $\pm$ 0.0 & 0.97 $\pm$ 0.15 & 1.0 $\pm$ 0.0 \\
\hline
Makespan ($\downarrow$) & 128.3 $\pm$ 40.87 & 72.37 $\pm$ 47.74 & 50.8 $\pm$ 28.40 & 58.45 $\pm$ 27.94 & 47.5 $\pm$ 25.28 & 55.6 $\pm$ 28.05 \\
\hline
\end{tabular}
\caption{Performance of each model checkpoint.}
\label{table:TrainingIterationPerformance}
\end{table*}

\subsection{Zero-Shot Deployment for Larger Team Sizes}
\label{subsec:explanation-quality-zero-shot}
One major benefit of GNN-based policies is the ability to directly deploy them to different team sizes, which has been done in works such as \cite{yu2022learning, hu2023graph, yang2023masp}. We analyze the explanation quality when deploying the trained policy on larger team sizes for the blind navigation task. This analysis can also be used to observe the explanation quality on out-of-distribution team sizes.

\subsubsection{Delta Fidelity}
As seen before, GraphMask does not gain any improvement in Delta Fidelity due to the sole reliance on hard binary masks and the approximately equal trade-off between Positive and Negative Fidelity that occurs when GraphMask is able to capture a larger subset of important agent-agent influences but still misses the other remaining important agent-agent influences. Also, both GNNExplainer and AttentionExplainer benefit from attention entropy minimization, with AttentionExplainer inhereting a larger improvement due to attention entropy minimization. However, note that, in general, as the test team size deviates further from the train team size (i.e. the deployed team size is more out-of-distribution), GNNExplainer performs more similarly to AttentionExplainer both in terms of fidelity and faithfulness. Thus, while GNNExplainer performs the best across models without attention entropy minimization, AttentionExplainer and GNNExplainer are interchangeably the best performers across the out-of-distribution evaluations. This can likely be due to the fact that while GNNExplainer has a difficult optimization to undertake and makes assumptions that work better on the graph datasets more common in the graph learning community, it is still not biased by the model itself, and treats the model as a black-box. On the other hand, AttentionExplainer is based solely on the attention values. Since the attention values are not trained to minimize the attention entropy on these out-of-distribution teams, even though there is evidently some entropy minimization effect occurring that transfers over to the out-of-distribution teams, it is likely less pronounced. Since the effect of attention entropy minimization likely deteriorates with more out-of-distribution teams, AttentionExplainer becomes less superior to GNNExplainer in terms of fidelity. 

\subsubsection{Unfaithfulness}
As seen before, GraphMask improves in faithfulness when the model is trained with attention entropy minimization, which is likely due to attention entropy minimization making the distribution sparser, resulting in agent-agent influences that are closer to a subgraph composed of only a binary adjacency matrix. Still, GraphMask does not, on average, achieve as faithful of explanations as GNNExplainer and AttentionExplainer. This is once again attributed to the limited representation capacity of GraphMask subgraphs since they are constrained to be hard binary masks, and attention entropy minimization can only ameliorate this, to an extent. In addition, we observe that the faithfulness of AttentionExplainer improves due to attention entropy minimization, while GNNExplainer has mixed improvement results in faithfulness due to attention entropy minimization. GNNExplainer performs the best across models without attention entropy minimization. With models trained with attention entropy, Attention Explainer performs better when the team is less out-of-distribution (i.e., the number of agents of the test team is closer to the number of agents of the train team). As the team size grows more out-of-distribution, the performance of AttentionExplainer and GNNExplainer becomes more interchangeable. This can likely be attributed to the fact that the model has not directly optimized for attention entropy minimization on these out-of-distribution team sizes, so the effect of attention entropy minimization deteriorates the more out-of-distribution the team size is. On the other hand, GNNExplainer treats the model as a black-box model and is not directly influenced by the attention values, but still has to solve a fairly difficult optimization problem and relies on assumptions that are more useful in the graph datasets that are more common in the graph learning community.

\subsubsection{Task Performance}
\tableautorefname~\ref{table:ZeroShotPerformance} presents the task performance metrics for each zero-shot train-test combination. Overall, the performance of the models trained with attention entropy minimization ranges from being comparable to outperforming the models trained without attention entropy minimization.

\begin{figure}[!h]
    \centering
    \includegraphics[width=0.33\linewidth]{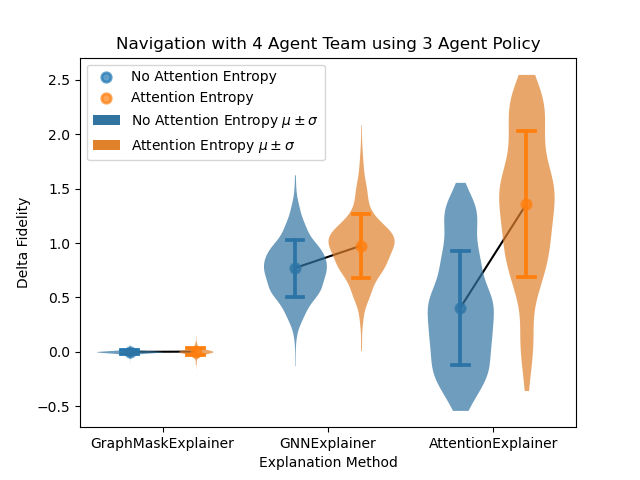}
    \includegraphics[width=0.33\linewidth]{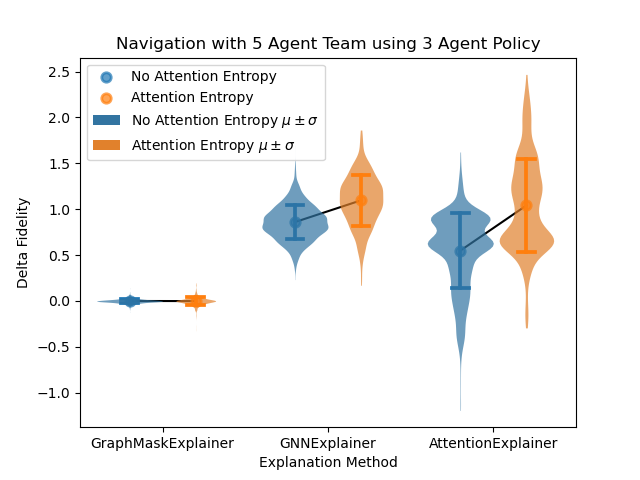}
    \includegraphics[width=0.33\linewidth]{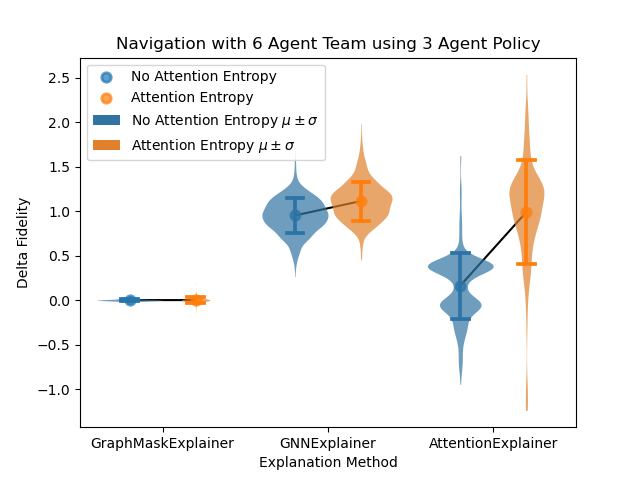}
    \includegraphics[width=0.33\linewidth]{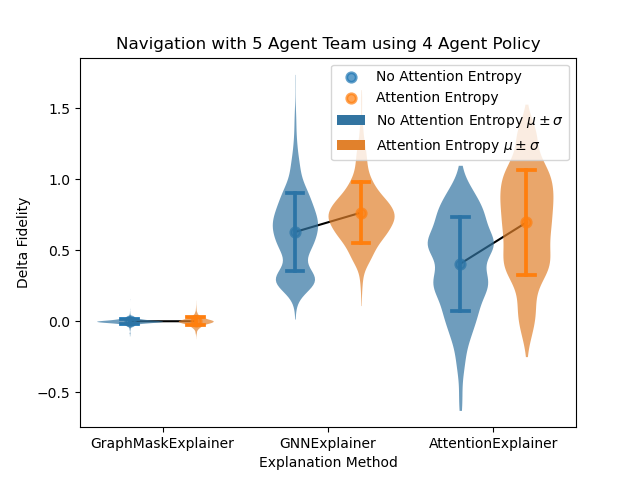}
    \includegraphics[width=0.33\linewidth]{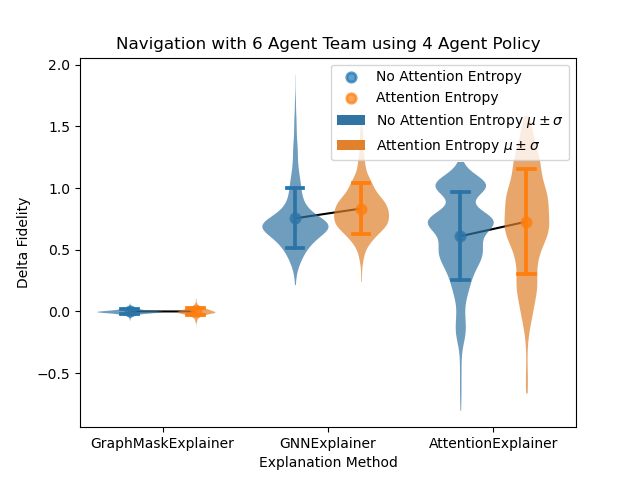}
    \includegraphics[width=0.33\linewidth]{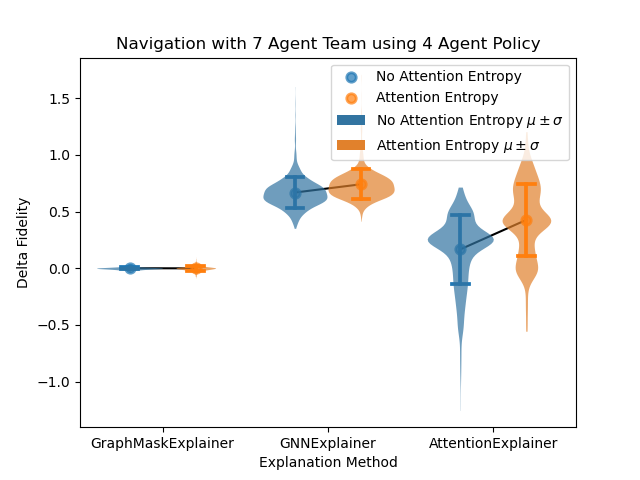}
    \includegraphics[width=0.33\linewidth]{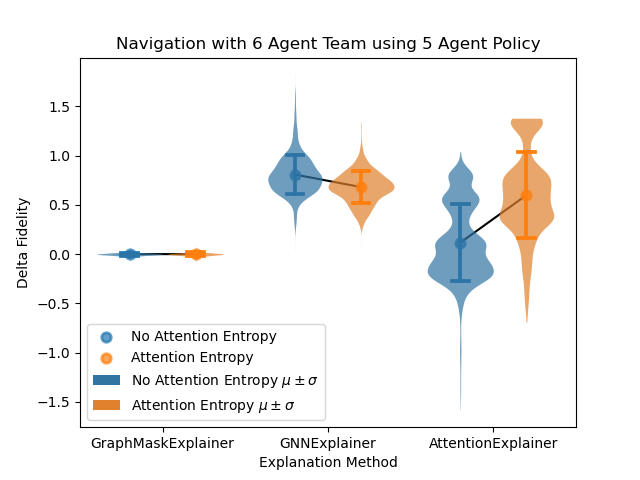}
    \includegraphics[width=0.33\linewidth]{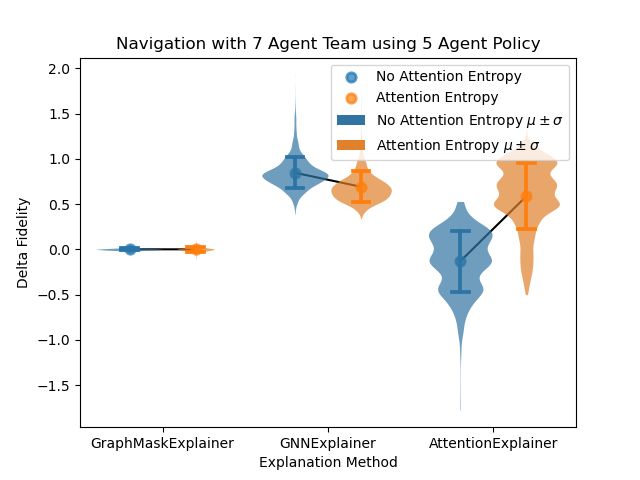}
    \includegraphics[width=0.33\linewidth]{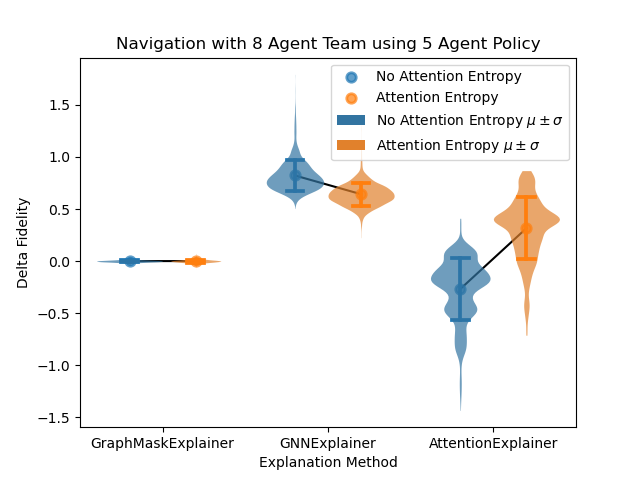}
    \caption{Delta Fidelity ($\uparrow$) of explanations generated by three explanation methods with and without proposed regularization. Each row represents the team size the policy was originally trained on (Top = 3, Middle = 4, Bottom = 5). Each column represents the larger team size the policy was deployed on (Left = +1 team size, Middle = +2 team size, Right = +3 team size)}
    \label{fig:explanationQualityOOD}
\end{figure}

\begin{figure}[!h]
    \centering
    \includegraphics[width=0.33\linewidth]{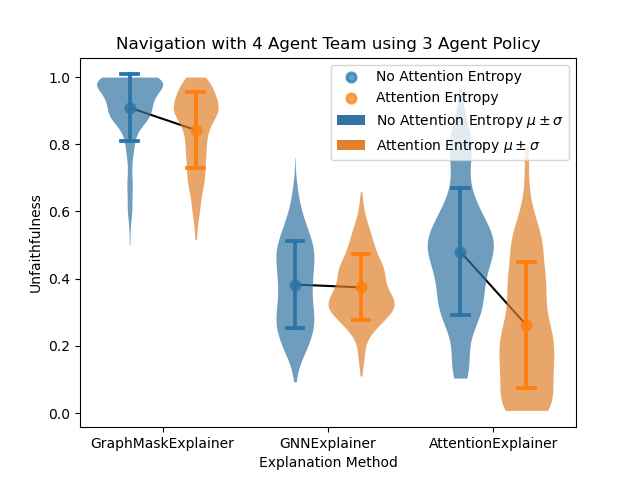}
    \includegraphics[width=0.33\linewidth]{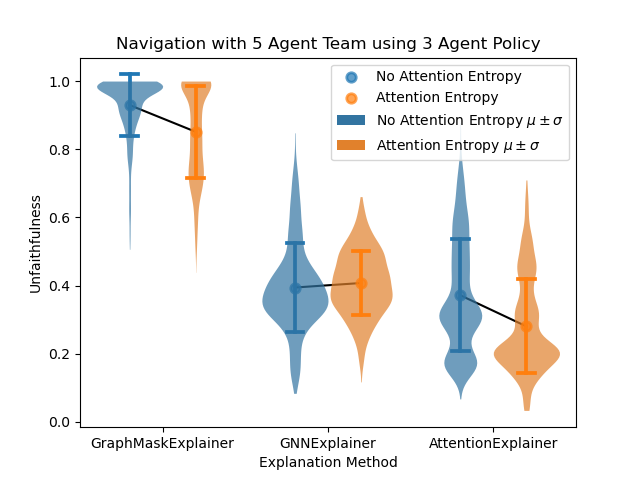}
    \includegraphics[width=0.33\linewidth]{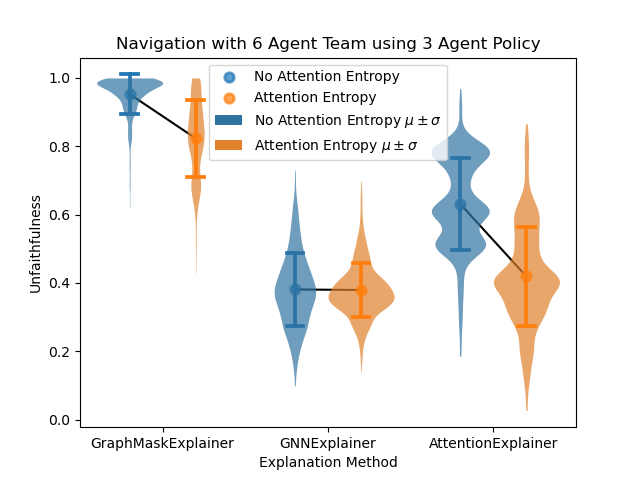}
    \includegraphics[width=0.33\linewidth]{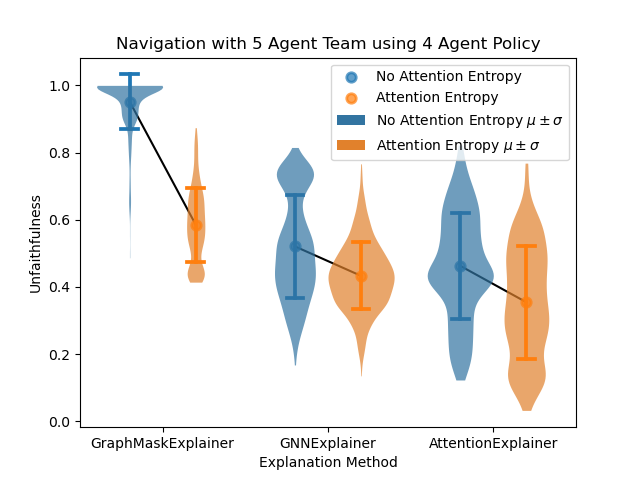}
    \includegraphics[width=0.33\linewidth]{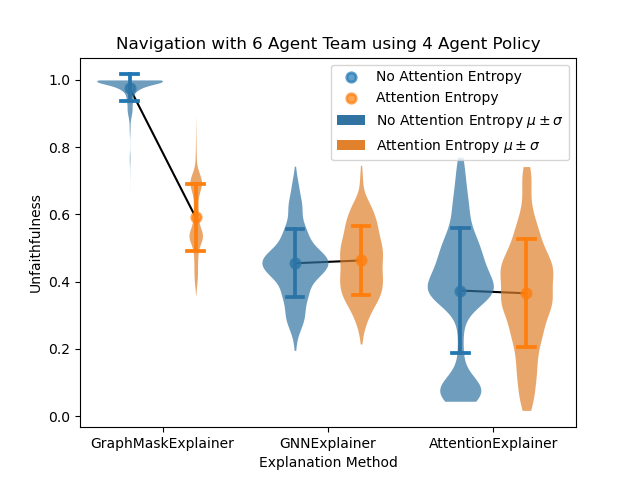}
    \includegraphics[width=0.33\linewidth]{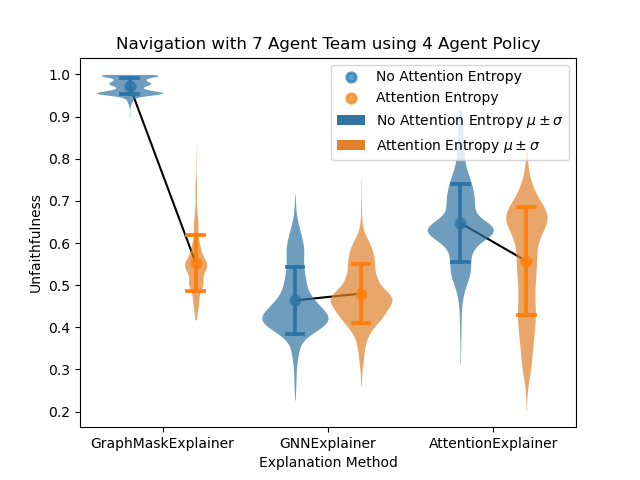}
    \includegraphics[width=0.33\linewidth]{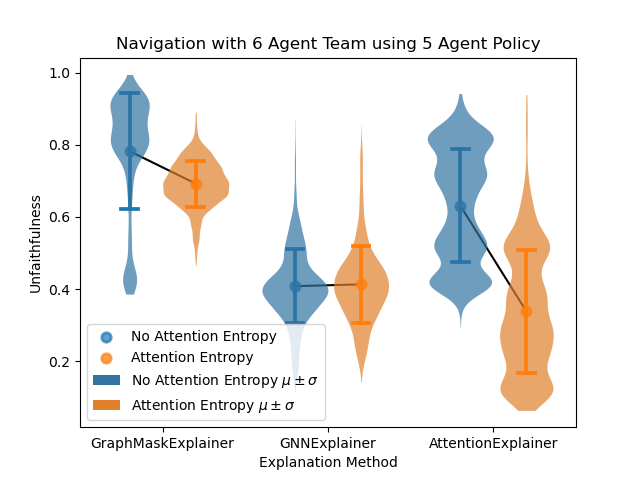}
    \includegraphics[width=0.33\linewidth]{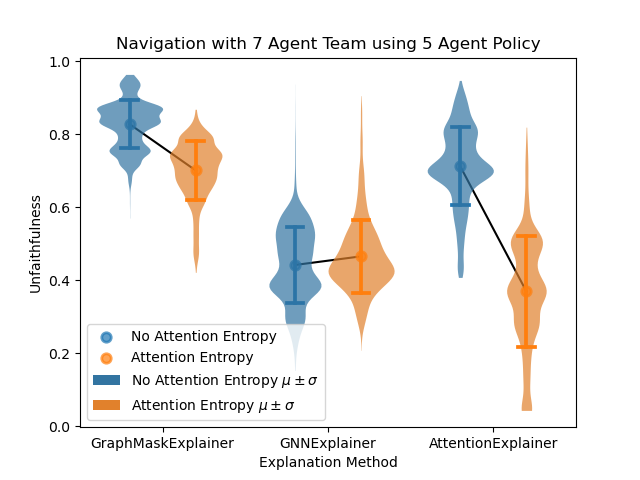}
    \includegraphics[width=0.33\linewidth]{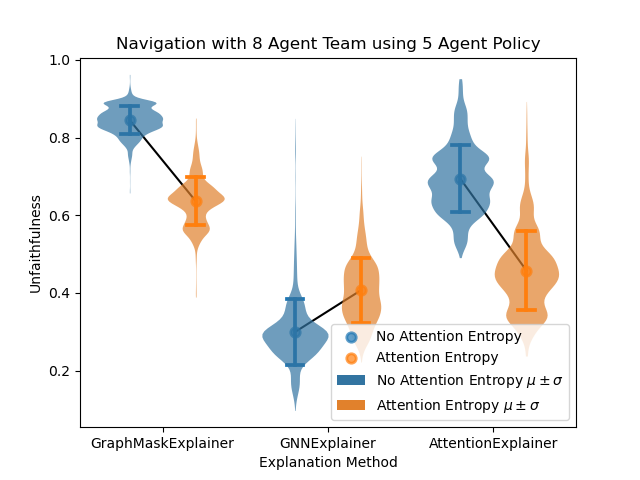}
    \caption{Unfaithfulness ($\downarrow$) of explanations generated by three explanation methods with and without proposed regularization. Each row represents the team size the policy was originally trained on (Top = 3, Middle = 4, Bottom = 5). Each column represents the larger team size the policy was deployed on (Left = +1 team size, Middle = +2 team size, Right = +3 team size)}
    \label{fig:explanationQualityOOD}
\end{figure}

\begin{table*}[bt]
\scriptsize
\centering
\begin{tabular}{|c|c|c|c|c|c|c|}
\hline
\multicolumn{7}{|c|}{\textbf{Blind Navigation Trained on N=3}} \\
\hline
\multirow{2}{*}{Performance Metric} & \multicolumn{2}{c|}{Tested on N=4} & \multicolumn{2}{c|}{Tested on N=5} & \multicolumn{2}{c|}{Tested on N=6} \\
    \cline{2-7}
     &  No Att. Ent. Min. & Att. Ent. Min. & No Att. Ent. Min. & Att. Ent. Min. & No Att. Ent. Min. & Att. Ent. Min. \\
\hline
Reward ($\uparrow$) & 2.68 $\pm$ 1.28 & 2.79 $\pm$ 0.69 & 4.74 $\pm$ 1.21 & 5.02 $\pm$ 0.99 & 3.98 $\pm$ 1.26 & 4.08 $\pm$ 1.68 \\
\hline
Success Rate ($\uparrow$) & 1.0 $\pm$ 0.0 & 0.93 $\pm$ 0.26 & 0.82 $\pm$ 0.37 & 0.95 $\pm$ 0.21 & 0.9 $\pm$ 0.29 & 0.88 $\pm$ 0.33 \\
\hline
No Agent Coll. Rate ($\uparrow$) & 0.93 $\pm$ 0.26 & 0.93 $\pm$ 0.26 & 0.92 $\pm$ 0.26 & 0.97 $\pm$ 0.15 & 0.67 $\pm$ 0.47 & 0.75 $\pm$ 0.43 \\
\hline
Makespan ($\downarrow$) & 61.57 $\pm$ 25.16 & 60.82 $\pm$ 33.83 & 154.57 $\pm$ 121.44 & 120.5 $\pm$ 75.15 & 116.6 $\pm$ 103.4 & 129.4 $\pm$ 110.33 \\
\hline
\hline
\multicolumn{7}{|c|}{\textbf{Blind Navigation Trained on N=4}} \\
\hline
\multirow{2}{*}{Performance Metric} & \multicolumn{2}{c|}{Tested on N=5} & \multicolumn{2}{c|}{Tested on N=6} & \multicolumn{2}{c|}{Tested on N=7} \\
    \cline{2-7}
     &  No Att. Ent. Min. & Att. Ent. Min. & No Att. Ent. Min. & Att. Ent. Min. & No Att. Ent. Min. & Att. Ent. Min. \\
\hline
Reward ($\uparrow$) & 5.08 $\pm$ 0.93 & 5.03 $\pm$ 0.70 & 6.55 $\pm$ 1.52 & 6.33 $\pm$ 1.62 & 5.14 $\pm$ 0.95 & 5.12 $\pm$ 0.99 \\
\hline
Success Rate ($\uparrow$) & 0.92 $\pm$ 0.26 & 1.0 $\pm$ 0.0 & 0.87 $\pm$ 0.33 & 0.95 $\pm$ 0.21 & 0.80 $\pm$ 0.40 & 0.80 $\pm$ 0.40 \\
\hline
No Agent Coll. Rate ($\uparrow$) & 0.92 $\pm$ 0.26 & 0.98 $\pm$ 0.15 & 0.8 $\pm$ 0.40 & 0.88 $\pm$ 0.33 & 0.85 $\pm$ 0.35 & 0.75 $\pm$ 0.43 \\
\hline
Makespan ($\downarrow$) & 125.17 $\pm$ 88.26 & 103.87 $\pm$ 58.46 & 149.82 $\pm$ 102.37 & 135.32 $\pm$ 82.67 & 158.62 $\pm$ 125.71 & 174.47 $\pm$ 130.42 \\
\hline
\hline
\multicolumn{7}{|c|}{\textbf{Blind Navigation Trained on N=5}} \\
\hline
\multirow{2}{*}{Performance Metric} & \multicolumn{2}{c|}{Tested on N=6} & \multicolumn{2}{c|}{Tested on N=7} & \multicolumn{2}{c|}{Tested on N=8} \\
    \cline{2-7}
     &  No Att. Ent. Min. & Att. Ent. Min. & No Att. Ent. Min. & Att. Ent. Min. & No Att. Ent. Min. & Att. Ent. Min. \\
\hline
Reward ($\uparrow$) & 6.03 $\pm$ 2.00 & 6.41 $\pm$ 1.61 & 7.33 $\pm$ 3.18 & 8.70 $\pm$ 2.30 & 5.04 $\pm$ 4.22 & 5.82 $\pm$ 3.48 \\
\hline
Success Rate ($\uparrow$) & 0.52 $\pm$ 0.49 & 0.72 $\pm$ 0.44 & 0.42 $\pm$ 0.49 & 0.52 $\pm$ 0.49 & 0.2 $\pm$ 0.40 & 0.45 $\pm$ 0.49 \\
\hline
No Agent Coll. Rate ($\uparrow$) & 0.87 $\pm$ 0.33 & 0.95 $\pm$ 0.21 & 0.6 $\pm$ 0.48 & 0.85 $\pm$ 0.35 & 0.52 $\pm$ 0.49 & 0.72 $\pm$ 0.44 \\
\hline
Makespan ($\downarrow$) & 282.7 $\pm$ 123.71 & 248.17 $\pm$ 114.98 & 333.72 $\pm$ 90.45 & 296.85 $\pm$ 108.79 & 362.7 $\pm$ 80.47 & 341.57 $\pm$ 80.20 \\
\hline
\end{tabular}
\caption{Performance of zero-shot deployment of policies trained on smaller team size on larger size test teams.}
\label{table:ZeroShotPerformance}
\end{table*}

\subsection{Limitations}
In our empirical analysis, we attempted to cover a wide variety of potential factors by considering multiple tasks in multi-robot coordination and architectural hyperparameters associated with multi-robot coordination. However, there are certain aspects not included in this work that can be further studied in future work. The first is other learning paradigms, such as imitation learning. The current study only focuses on reinforcement learning as finding an expert or demonstration dataset for multi-agent systems is challenging, especially for the chosen tasks. The second is alternative neural architecture paradigms. We focus on the most representative structure of encoder-GNN-decoder, but we focused mostly on MLPs as opposed to other types of neural architectures. Finally, there are other potential ways to generate sparse attention or edge weights besides attention entropy minimization. Especially for GraphMask, one could try sampling the attention weights as a Bernoulli trial to generate hard edge masks, which could be more in-distribution for GraphMask. These are all left for future work and study.

\end{document}